\documentclass[11pt]{article}

\usepackage{amssymb,amsthm,amsmath}
\usepackage{libertine}
\usepackage[libertine]{newtxmath}
\usepackage[lined,boxed,ruled,norelsize,algo2e,linesnumbered]{algorithm2e}


\usepackage{subfig}
\usepackage[usenames,dvipsnames]{xcolor}
\usepackage[colorlinks,citecolor=blue,linkcolor=BrickRed]{hyperref}\usepackage[colorlinks,citecolor=blue,linkcolor=BrickRed]{hyperref}
\usepackage{fullpage}
\usepackage{graphicx}
\usepackage{xspace}
\usepackage{enumitem}
\usepackage{thmtools,thm-restate}
\usepackage{wrapfig}
\usepackage{algpseudocode}

\graphicspath{{figures/}}

\newtheorem{theorem}{Theorem}

\newtheorem{lemma}[theorem]{Lemma}

\newtheorem{corollary}[theorem]{Corollary}

\newtheorem{claim}[theorem]{Claim}

\newif\ifFULL
\FULLtrue

\newcounter{note}[section]
\newcommand{\sahil}[1]{\refstepcounter{note}$\ll${\bf Sahil~\thenote:} 
  {\sf \color{blue}  #1}$\gg$\marginpar{\tiny\bf   SS~\thenote}} 
  
\newcommand{\haotian}[1]{\refstepcounter{note}$\ll${\bf Haotian~\thenote:}
  {\sf \color{red}  #1}$\gg$\marginpar{\tiny\bf HJ~\thenote}}

\newcommand{\depot}{\ensuremath{\mathsf{root}}\xspace}

\newcommand{\p}{\ensuremath{\mathbb{P}}}

\newcommand{\E}{\mathbb{E}}

\newcommand{\OPT}{\ensuremath{\mathsf{OPT}}\xspace}
\newcommand{\ALG}{\ensuremath{\mathsf{ALG}}\xspace}

\newcommand{\one}{\textsf{1}\xspace}

\newcommand{\eat}[1]{}
\newcommand{\hide}[1]{{\Large \color{red} Contents here are hidden! To reveal contents, remove this command.}}

\newcommand{\polylog}{\ensuremath{\textsf{polylog}}\xspace}
\newcommand{\ALGori}{\ensuremath{\ALG_{\textsf{Bicrit-Orient}}}\xspace}
\newcommand{\ALGaux}{\ensuremath{\ALG_{\textsf{Rep}}}\xspace}
\newcommand{\ALGDP}{\ensuremath{\ALG_{\textsf{DP}}}\xspace}
\newcommand{\ALGO}{\ensuremath{\ALG_{\textsf{Orient}}}\xspace}
\newcommand{\OPTO}{\ensuremath{\OPT_{\textsf{Orient}}}\xspace}
\newcommand{\ADAP}{\ensuremath{\mathsf{ADAP}}\xspace}
\newcommand{\ALGmeta}{\ensuremath{\ALG_{\textsf{Meta}}}\xspace}
\newcommand{\ALGkTSP}{\ensuremath{\ALG_{\KTSP}}\xspace}
\newcommand{\AdkTSP}{\ensuremath{\ALG_{Ad-kTSP}}\xspace}

\newcommand{\KTSP}{\ensuremath{\textsf{$k$-TSP}}\xspace}
\newcommand{\KMST}{\ensuremath{\textsf{$k$-MST}}\xspace}

\newcommand{\Orient}{\ensuremath{\textsf{Orienteering}}\xspace}

\newcommand{\ori}{\ensuremath{\textsf{Ori}}\xspace}
\newcommand{\StochKTSP}{\ensuremath{\textsf{Stoch-Reward $k$-TSP}}\xspace}
\newcommand{\StochKnap}{\ensuremath{\textsf{Stoch-Knapsack Cover}}\xspace}
\newcommand{\StochKCost}{\ensuremath{\textsf{Stoch-Cost $k$-TSP}}\xspace}

\newcommand{\poly}{\ensuremath{\mathrm{poly}}\xspace}
\newcommand{\ALGStochKTSP}{\ensuremath{\mathsf{\ALG_{\textsf{Stoch-Reward}}}}\xspace}
\newcommand{\ALGStochKCost}{\ensuremath{\mathsf{\ALG_{\textsf{Stoch-Cost}}}}\xspace}

\newcommand{\old}{\mathsf{old}}
\newcommand{\new}{\mathsf{new}}

\newcommand{\jcrit}{\ensuremath{{j_{\textsf{crit}}}}}
\newcommand{\tildejcrit}{\ensuremath{{\widetilde{j}_{\textsf{crit}}}}}
\newcommand{\tildeY}{\ensuremath{\widetilde{Y}}}

\newcommand{\IGNORE}[1]{}

\newenvironment{proofof}[1]{\smallskip\noindent{\bf Proof of #1.}}%
        {\hspace*{\fill}$\Box$\par}

\title{ 
Algorithms and Adaptivity Gaps for Stochastic $k$-TSP}

\author{
Haotian Jiang
\thanks{Paul G. Allen School of Computer Science \& Engineering, University of Washington. Email:  \texttt{jhtdavid@cs.washington.edu.}}
\and Jian Li
\thanks{Institute for Interdisciplinary Information Sciences, Tsinghua University. Email:\texttt{lijian83@mail.tsinghua.edu.cn.}}
\and Daogao Liu
\thanks{Department of Physics, Tsinghua University. Email: \texttt{liudg16@mails.tsinghua.edu.cn.}}
\and Sahil Singla
\thanks{Computer Science Department at Princeton University and School of Mathematics at Institute for Advanced Study. Supported in part by the Schmidt Foundation. Email: \texttt{singla@cs.princeton.edu}.}
}

\date{\today }

\begin{document}
\maketitle
\thispagestyle{empty}

\begin{abstract}
Given a metric $(V,d)$ and a $\depot \in V$, the classic $\KTSP$  problem is to find a tour originating at the $\depot$ of minimum length that visits at least $k$ nodes in $V$. In this work, motivated by applications where the input to an optimization problem is uncertain,  we study two stochastic versions of  $\KTSP$.

In \StochKTSP, originally defined by Ene-Nagarajan-Saket~\cite{ENS17},  each vertex $v$ in the given metric $(V,d)$ contains a stochastic reward $R_v$. The goal  is to adaptively find a tour  of minimum \emph{expected} length that collects at least  reward $k$; here ``adaptively'' means our next decision may depend on previous outcomes. Ene et al. give an $O(\log k)$-approximation adaptive algorithm for this problem, and left open if there is an $O(1)$-approximation algorithm. We totally resolve their open question, and even give an $O(1)$-approximation \emph{non-adaptive} algorithm for \StochKTSP. 

We also introduce and obtain similar results  for the  \StochKCost problem. In this problem each vertex $v$ has a stochastic cost $C_v$, and the goal is to visit and select at least $k$ vertices to minimize the expected \emph{sum} of tour length and cost of selected vertices. Besides being a natural stochastic generalization of \KTSP, this problem is also interesting because it generalizes the  Price of Information framework~\cite{Singla-SODA18} from deterministic probing costs to metric probing costs.

Our techniques are based on two crucial ideas: ``repetitions'' and ``critical scaling''. In general, replacing a random variable with its expectation  leads to very poor results. We show that for our problems, if we truncate the random variables at an ideal threshold, then their expected values form a good surrogate.
Here, we rely on running several repetitions of our algorithm with the same threshold, and then argue concentration using Freedman's and Jogdeo-Samuels' inequalities. 
Unfortunately, this ideal threshold depends on how far we are from achieving our target $k$, which a non-adaptive algorithm does not know. To overcome this barrier, we truncate the random variables at various different scales and identify a ``critical'' scale.

\end{abstract}
\newpage


\setcounter{page}{1}


\section{Introduction}

Consider a scenario where a salesperson must sell some quota of brushes in order to win a trip to Hawaii. The salesperson knows the time it takes to travel between different cities and the demand at each city. What is the best route to take to sell the quota while spending the least amount of time? This exact scenario was  described by  Awerbuch et al.~\cite{AwerbuchABV-SICOMP98} to motivate the study of TSP problems where the algorithm has to also decide which cities to visit. 
A cleaner version of this problem, first introduced by Ravi et al.~\cite{ravi1996spanning}, is the $\KTSP$ problem where we assume that each city has a unit demand. 
Formally,  given a metric $(V,d)$ with a  $ \depot \in V$ and a \emph{target} $k \in \mathbb{Z}_{\geq 0}$, the $\KTSP$ problem is to find a tour  that originates at the $\depot$ and visits at least $k$ vertices, while minimizing the total travel time. There is a long line of work trying to design better approximation algorithms for the $\KTSP$ problem~\cite{AwerbuchABV-SICOMP98,rajagopalan1995logarithmic,BlumRV-STOC96,Garg1996,arya1998,arora2000}, and the state-of-the-art is a $2$-approximation algorithm due to Garg~\cite{Garg2005}\footnote{A closely-related variant is called the $\KMST$ problem. Both problems are equivalent up to a constant approximation factor.}. 

In this work we consider stochastic versions of the  \KTSP problem: what if the  salesperson does not  know the exact demand in each city, or what if the salesperson need to  spend some  uncertain time in each city to complete the city's demand?
Indeed, there is a long line of work studying classical optimization problems  where we begin only with estimates (probability distributions) on the  input parameters. The algorithm has to \emph{adaptively probe} parameters (inspect elements) by paying some ``cost'' before realizing their exact values; here ``adaptively'' means that our decisions may depend on the outcomes of already probed elements. Such stochastic probing problems have been  well-studied in both  maximization and  minimization settings~\cite{DGV-FOCS04,GoemansV-LATIN06,GuhaM09,GKMR-FOCS11,GM-SODA07TALG12,GKNR-SODA12,BGLMNR-Algorithmica12,GN-IPCO13,BN-IPCO14,Ma-SODA14,AN16,GNS-SODA17,ENS17,FuLX18,Singla-SODA18,BSZ-RANDOM19,GuptaJSS-IPCO19}.

There are two natural ways of defining the stochastic \KTSP problem, depending on the type of input uncertainty. 
In the \StochKTSP problem, first introduced by Ene-Nagarajan-Saket~\cite{ENS17}, we
incorporate uncertainty in the vertex demands. Formally, we   
assume that the demand at each vertex is drawn independently from a known distribution, and the goal is to \emph{adaptively} find a tour $\Pi$ that obtains the target demand $k$, 
while minimizing the total \emph{expected} travel time\footnote{Our \StochKTSP  problem is called the ``Stochastic $k$-TSP'' problem in~\cite{ENS17}. We rename it to differentiate it from \StochKCost.}.

We also study  \StochKCost  where each city still has a unit demand, but the salesperson will have to spend an additional \emph{completion} time (drawn from a known distribution) at each vertex before meeting its unit demand.
The goal is to  {adaptively} find a tour $\Pi$ that completes the target demand $k$, while minimizing the {expected} \emph{sum} of total  travel and completion times. Notice, our algorithm  finds the exact completion time of a vertex only after visiting it, and may choose not to complete it (i.e., not meet its unit demand) if the completion time seems too long. This idea of studying stochastic completion times at  vertices is not new, and has been previously used  in the study of stochastic \Orient problems, which in some sense is the dual to our \StochKCost problem~\cite{GKNR-SODA12,BN-IPCO14}.

A common theme in the study of stochastic probing problems is to understand the power of adaptivity. Indeed, while the optimal algorithms can  fully adapt to the outcomes, and hence may not even have a polynomial-size representation,  a \emph{non-adaptive} algorithm makes all its decisions upfront independent of the observed outcomes (except perhaps the stopping time). Being non-adaptive has several benefits like they are  simpler to find, easily parallelizable, and have a  poly-size representation. So ideally for a probing problem we would like to design non-adaptive algorithms  with performance close to the optimal adaptive algorithms, or in other words design non-adaptive algorithms with a small \emph{adaptivity gap}. The main results of our work is to show that both the above \StochKTSP and \StochKCost  problems have a constant adaptivity-gap. That is, there exist  fixed-tours starting at the \depot which the algorithm can take until it obtains the target demand $k$, which guarantee an expected total time at most a constant factor more than the expected total time of the optimal adaptive tour. Moreover, for  distributions with polynomial support, we give poly-time algorithms to find such tours.

Our constant adaptivity-gap for \StochKTSP answers the main open question of~\cite{ENS17}, who showed an $O(\log^2 k)$ bound on the adaptivity gap. The constant adaptivity gap result for \StochKCost might also seem surprising because it is known that the related Stochastic \Orient problem has a super-constant adaptivity gap~\cite{BN-IPCO14}.

In the rest of this section we first formally state our problems and results, and then discuss our high-level techniques and other related work.

\subsection{\StochKTSP}

The following \StochKTSP was first defined by Ene et al.~\cite{ENS17}. 

\noindent \textbf{\StochKTSP:} We are given a metric $(V, d)$ with a $\depot \in V$ and each vertex $v \in V$ has an independent integral\footnote{This is without loss of generality. Our result also generalizes to the case where rewards are real numbers via re-scaling.} stochastic\footnote{We assume that the distribution is discrete and is given explicitly.} reward $R_v \in \mathbb{Z}_{\geq 0}$. All reward distributions are given as input but the actual reward instantiation $R_v$ is only known when the algorithm visits vertex $v$. Given a target value $k \in \mathbb{Z}_{\geq 0}$, the goal is to adaptively find a tour $\Pi$ originating at $\depot$ that collects at least $k$ reward (i.e., $\sum_{v \in \Pi} R_v \geq k$) while  minimizing the expected tour length. 

This problem captures several well-studied problems; e.g., it captures the \StochKnap problem where the metric $(V,d)$ is a weighted star: given a target $k$ and $n$ items where item $i\in [n]$ has both a deterministic cost $C_i \in \mathbb{R}_{\geq 0}$ and an independent stochastic reward $R_i \in \mathbb{Z}_{\geq 0}$, the \StochKnap problem is to adaptively obtain a total reward of at least $k$ at the minimum expected cost.
This problem was studied  by Deshpande et al.~\cite{DeshpandeHK-TALG16}, 
and they gave
an \emph{adaptive} 2-approximation algorithm. However, even in this special case, 
it was not know if there is a 
non-adaptive
constant factor 
approximation algorithm. 

The first non-trivial results for the \StochKTSP problem were obtained by Ene et al.~\cite{ENS17}. They gave an $O(\log^2 k)$-approximation non-adaptive algorithm and an $O(\log k)$-approximation adaptive algorithm.
On the hardness side, however, they only gave a lower bound of $e$ on the adaptivity gap. This left open closing the wide gap on the adaptivity gap for \StochKTSP.
We resolve their open question by giving     a non-adaptive $O(1)$-approximation algorithm. 

\begin{restatable}{theorem}{stochkTSP}
\label{thm:StochKTSP}
There is a non-adaptive $O(1)$-approximation algorithm for the $\StochKTSP$ problem.
\end{restatable}

The difficulty in \StochKTSP arises because the expected reward is a poor indicator of how much we care about a node. An extreme example is  a vertex with a large expected reward, but which is non-zero with nearly zero probability.  
It is therefore reasonable to truncate the reward distributions at the remaining target reward.
However, it is not clear why such an approach would work, and moreover this approach is adaptive as it  depends on the remaining target.


\subsection{\StochKCost}

We formally define the \StochKCost problem. 

\noindent \textbf{\StochKCost:}  We are given a metric $(V, d)$ with a $\depot \in V$ and each vertex $v \in V$ has an independent stochastic cost $C_v \in \mathbb{R}_{\geq 0}$. All cost distributions are given as input but the actual cost instantiation $C_v$ is only known when vertex $v$ is visited. Suppose a vertex $v$ can only be \emph{selected} if:  (1) $v$ is visited and (2) we are currently\footnote{\label{footnote:DistSelect} Up to a factor of 2, this version of the problem is equivalent to the version where restriction (2) is removed.} at vertex $v$.
The goal is to  adaptively find a tour $\Pi$ originating at $\depot$ that selects a set $S$ of $k$ visited vertices while minimizing the expected {\em total} cost, which is the sum of the tour length and the cost of the selected vertices:
\[
\E \Big[d(\Pi) + \sum_{v \in S} C_v \Big].
\]

Apart from being a natural generalization of the classical \KTSP problem,  \StochKCost
is also motivated due to its connections to price of information~\cite{Singla-SODA18}. In particular, it generalizes  the \textsf{Minimization $k$-Pandora's Box} problem studied in~\cite{KWW-EC16,Singla-SODA18}.
In this problem we are given a target $k$ and $n$ items, where each item $i \in [n]$ has a known probing price $\pi_i \in \mathbb{R}_{\geq 0}$ and an  independent stochastic cost $C_i$.
The exact cost $C_i$ is only revealed after we pay price $\pi_i$. The goal is to adaptively probe and select $k$ of the probed items to minimize the expected total selection cost plus probing price. \StochKCost captures this problem on a star metric where node $i$ is at a distance $\pi_i/2$ from the $\depot$. Thus, we can view the \StochKCost problem as generalizing the price of information framework to a \emph{metric} setting, where the price of probing is not fixed but given by a general metric.
Our next result gives a non-adaptive $O(1)$-approximation algorithm for  \StochKCost.

\begin{restatable}{theorem}{stochkCost}
\label{thm:StochKCost}
There is a non-adaptive $O(1)$-approximation algorithm  for the $\StochKCost$ problem.
\end{restatable}

Our main techniques in the proof of Theorem~\ref{thm:StochKCost} are similar to those for \StochKTSP. In fact, in Section~\ref{sec:OurApproach} we present  a generic framework that can be used to solve both these problems. It will be interesting to find other applications of our framework in future work.  

\subsection{High-level Techniques}
\label{sec:highleveltechnique}

We assume that after re-scaling, the  distance between any pair of vertices is at least $1$. 

\smallskip
\noindent \textbf{\StochKTSP.} A standard idea in the design of approximation algorithms on a metric is to operate in phases, where in  phase~$i$ our algorithm is allowed a budget of $2^i$. 
Intuitively, this corresponds to the algorithm imagining that the optimal adaptive tour has length $\Theta(2^i)$.
In each phase, a na\"{\i}ve algorithm would be to collect as much {\em expected} reward as possible within budget $2^i$ (say, by solving an instance of the \Orient problem).
However, in general the performance of such a na\"{\i}ve algorithm can be arbitrarily bad. 
E.g., suppose $k-\sqrt{k}$ reward is easy to get, now for the remaining reward the na\"{\i}ve algorithm  prefers a vertex having a reward of $k$ with probability $10/\sqrt{k}$ and $0$ otherwise, as opposed to a vertex with a deterministic reward of $\sqrt{k}$ (assuming both are at the same distance). 

A natural fix to the above issue is to  {\em truncate} the reward distributions  at the \emph{remaining} target reward  and then take  expectations.
Not only is this algorithm {\em adaptive}, it is not even clear why it would work.
Indeed, Ene et al.~\cite{ENS17} give an example (see Example 2 in~\cite{ENS17}) where this algorithm has an $\Omega(\log k)$-approximation factor.
Our first idea is to run $O(1)$ \emph{repetitions} of a bi-criteria \Orient algorithm using the {\em same} truncation (i.e., the initial remaining target) for all these repetitions.
Our analysis applies Freedman's~\cite{freedman1975tail} and Jogdeo-Samuels~\cite{JS68} inequalities to argue concentration, and relies crucially on not updating the remaining target for these $O(1)$ repetitions.

\IGNORE{
To get an adaptive algorithm, the following ``repetition'' idea was used in~\cite{ENS17}: repeatedly collect as much expected reward (truncated at current remaining target) as possible within budget $2^i$ (via an \Orient algorithm).
However, $\Theta(\log k)$ repetitions are needed for their results and this is why their algorithms have approximation factors of $\Omega(\log k)$. 
In fact, they also show that $\Omega(\log k)$ repetitions are necessary via a counter-example (which we include in Appendix~\ref{sec:logkRepeatExample} for completeness).
The first novelty of our approach lies in bringing the number of repetitions down to $O(1)$ by using the {\em same} truncation (which is the remaining target in the first repetition) throughout the $O(1)$ repetitions of a {\em bi-criteria} \Orient algorithm (discussed in Section~\ref{subsec:OrientAlg}). 
Our analysis relies crucially on applying Freedman's~\cite{freedman1975tail} and Jogdeo-Samuels~\cite{JS68} inequalities to argue concentration.
}

Nevertheless, the above approach depends on the remaining target, which is unknown to a {\em non-adaptive} algorithm.
One  could bypass this  by  truncating the reward distributions at $\log k$  different {\em scales}, where scale $j$ corresponds to the remaining target being roughly $k/2^j$, applying the previous $O(1)$ repetitions idea at each scale, and visiting  the union of all tours. 
Unfortunately, this immediately loses a  $\log k$ approximation factor. 
Our second idea is  ``critical scaling''  in which we identify a ``critical'' scale $\jcrit$ among the $\log k$ possible scales, and only include tours for scales  $\jcrit-1$ and $\jcrit$.  This critical scale $\jcrit$ roughly (but not quite) corresponds to a ``phase transition'' from an underestimation to an overestimation of the remaining target. A priori it is not clear why such a ``critical'' scale can be found non-adaptively, but the  concentration properties of the above $O(1)$ repetitions allow us to find it efficiently.

\smallskip
\noindent \textbf{\StochKCost.}
We obtain our result for \StochKCost in a similar way. 
An immediate challenge, however,  is how should we truncate the cost distributions $C_v$, say even if the remaining target $k'$ is known? 
One natural way  is by looking at $\p[C_v \leq O(2^i/k')]$, where $O(2^i/k')$ is the ``average'' cost per remaining reward  in phase $i$. 
But such an approach would fail when some vertices in the optimal tour have costs much smaller than $O(2^i/k')$, while the other vertices have much higher costs.
We overcome this by considering $\p[C_v \leq O(2^{i-j})]$ for all possible scales $j \in \{0,\cdots, i + \log n\}$, identifying a ``critical'' scale $\jcrit$, and again only including the tours for   scales $\jcrit-1$ and $\jcrit$. 
To identify the critical scale, we need to
evaluate the maximum target a tour at a given scale can get within cost budget $2^i$ with constant probability. We show this can be approximately computed  via dynamic programming.

\subsection{Further Related Work}

There is a long line of work studying the classic \KTSP and the related \KMST problem; we refer the readers to  Garg's beautiful $2$-approximation  paper and the references therein~\cite{Garg2005}. 

A formal study on the benefits of adaptivity for stochastic combinatorial optimization problems started with the seminal work of Dean et al.~\cite{DGV-FOCS04}. They showed that for the stochastic knapsack problem, where items  sizes are independently drawn and we need to fit them in a knapsack of size $B$, there is an $O(1)$-approximation non-adaptive algorithm. This factor was later improved to a $(2+\epsilon)$-approximation in~\cite{BGK-SODA11,Ma-SODA14}.
The minimization version of the stochastic knapsack
problem, which is known as the \StochKnap problem, 
was studied by Deshpande et al.~\cite{DeshpandeHK-TALG16}, and is a special 
case of \StochKTSP as we mentioned before.
The unbounded version of \StochKnap
(each item has infinite number of copies)
was studied by \cite{jiang2019fptas} and they provide
an FPTAS for this problem.

Gupta et al.~\cite{GKNR-SODA12} generalized the stochastic knapsack problem to the stochastic orienteering problem, where each stochastic item now resides on a vertex of a given metric, and we need to fit both the tour length and the item sizes inside our budget $B$. They give an $O(\log\log B)$-approximation non-adaptive algorithm for this problem. Bansal and Nagarajan~\cite{BN-IPCO14} later showed that this problem has  no constant-approximation non-adaptive algorithm. These works inspired Ene et al.~\cite{ENS17} to study \StochKTSP, a natural  minimization variant of the stochastic orienteering problem. Prior to our work, it was conceivable that this minimization problem also has a super-constant adaptivity gap, like stochastic orienteering.

Motivated by different applications that solve discrete problems under an uncertain input,  other related stochastic probing models have been studied. We refer the readers to Singla's Ph.D. Thesis for a  survey~\cite{Singla-Thesis18}. 
Of particular interest to us is the Price of Information model~\cite{Singla-SODA18}, which was inspired from the work on Pandora's box~\cite{Weitzman-Econ79,KWW-EC16}. Their \textsf{Minimization $k$-Pandora's Box} problem  inspired us to define \StochKCost,  which generalizes the probing costs from being fixed to being on a metric.  
Although an optimal strategy is known for \textsf{Minimization $k$-Pandora's Box}, the problem becomes APX-hard on a metric  as it generalizes \KTSP. 



\paragraph{Organization}
We start with some preliminary definitions and lemmas in Section~\ref{sec:preliminary}. 
In Section~\ref{sec:OurApproach}, we describe our general framework for both \StochKTSP and \StochKCost, and prove some key lemmas that will be used throughout the paper.
We give our non-adaptive $O(1)$-approximation algorithm for \StochKTSP that proves Theorem~\ref{thm:StochKTSP} in Section~\ref{sec:StochKTSP}. 
The non-adaptive $O(1)$-approximation algorithm for \StochKCost that proves Theorem~\ref{thm:StochKCost} is in Section~\ref{sec:StochKCost}.


\section{Preliminaries}
\label{sec:preliminary}

\subsection{Adaptive vs Non-Adaptive Algorithms}

Any feasible solution to our stochastic problems can be described by a \emph{decision tree}, where nodes correspond to vertices that are visited and branches correspond to instantiations of the observed random variables. Even if the degree of every vertex is a constant, the size of such decision trees can be exponentially large in its height. 
These solutions are called {\em adaptive} because the choice of the next vertex to visit depends on the outcomes of the already visited nodes.

We also consider the special class of {\em non-adaptive} solutions that is
described simply by an ordered list of vertices: the policy involves visiting vertices in the
given order until a certain stopping criterion is met. 
Such non-adaptive solutions are often preferred over adaptive solutions because they are easier to implement. 

In this work  we only study minimization problems. We compare the performance of our algorithm with that of the optimal {\em adaptive} algorithm, which is denoted by $\OPT$. We abuse notation and also use $\OPT$ to denote the expected objective of the optimal adaptive algorithm. 
 We say an algorithm is $\alpha$-approximation for $\alpha \geq 1$ if the expected objective of the algorithm is at most $\alpha \cdot \OPT$.
Ideally, we want to design non-adaptive algorithms whose performance is comparable to the optimal adaptive algorithm. Since this  is not always possible,  it 
is important to bound the {\em adaptivity gap}, which is the worst-case ratio between the expected objectives of the optimal non-adaptive and the optimal adaptive algorithms.

\subsection{Probability Inequalities}

Our proofs will require the following probability inequalities. 
We start with a bound on the median of independent Bernoulli random variables due to Jogdeo and Samuels~\cite{JS68}. 
Given $n$ independent Bernoulli random variables $X_1,\cdots, X_n$ where $X_i$ has success probability $p_i \in [0,1]$, let $X:=\sum_{i=1}^n X_i$ be their sum.
Define the median of $X$ to be any integer $m \in \mathbb{Z}_{\geq 0}$ such that $\min\left\{\p[X \geq m], \p[X \leq m] \right\} \geq 1/2$.

\begin{theorem}[Theorem 3.2 and Corollary 3.1~\cite{JS68}]
\label{thm:JogdeoSamuels}
Let $X= \sum_{i=1}^n X_i$ be the sum of $n$ independent Bernoulli random variables where $X_i$ has success probability $p_i \in [0,1]$.
If $\E[X]$ is an integer $k$, then the median of $X$ is also $k$. If $k < \E[X] < k+1$ for some integer $k$,
then the median of $X$ is either $k$ or $k+1$.
\end{theorem}

\IGNORE{
\begin{theorem}[Thm 3.2 and Cor 3.1~\cite{JS68}]
\label{thm:JogdeoSamuels}
If the mean number of successes in $n$ independent Bernoulli trials is an integer $k$, then the median is also $k$. If the mean number of successes in $n$ independent Bernoulli trials is between the integer $k$ and $k+1$ then the mean is either $k$ or $k+1$.
\end{theorem}}

We will also need the following martingale inequality due to Freedman~\cite{freedman1975tail}.

\begin{theorem}[Freedman's Inequality, Theorem 1.6 in~\cite{freedman1975tail}]
\label{thm:FreedmanInequality}
Consider a real-valued martingale sequence $\{X_t\}_{t\geq 0}$ such that $X_0=0$, and $\E[X_{t+1}|\mathcal{F}_t]=0$ for all $t$, where $\{\mathcal{F}_t\}_{t\geq 0}$ is the filtration defined by the martingale. Assume that the sequence is uniformly bounded, i.e., $|X_t|\leq M$ almost surely for all $t$. Now define the predictable quadratic variation process of the martingale to be $W_t=\sum_{j=1}^t \E[X_j^2|\mathcal{F}_{j-1}]$ for all $t\geq 1$. Then for all $\ell \geq 0$ and $\sigma^2>0$ and any stopping time $\tau$, we have
\[
\p\Big[ \Big|\sum_{j=0}^\tau X_j \Big|\geq \ell \wedge W_\tau \leq \sigma^2 \text{for some stopping time } \tau \Big] \leq 2\exp\Big(- \frac{\ell^2/2}{\sigma^2+M \ell/3} \Big).
\]
\end{theorem}

\begin{theorem}
[Chernoff Bound]
\label{thm:ChernoffBound}
Let $X_1,X_2,\cdots,X_n$ be independent random variables taking values in $[0,1]$ and define $X := \sum_{i \in [n]} X_i$. Then for any $\delta\in[0,1]$, we have
\[
\p\left[X \leq (1-\delta) \cdot \E[X] \right]\leq \exp\left(-\delta^2\cdot \E[X]/2 \right).
\]

\end{theorem}

\subsection{A Bi-Criteria Algorithm $\ALGori$ for \Orient}
\label{subsec:OrientAlg}

We formally define the  well-known \Orient problem. 

\noindent \textbf{\Orient:} Given a metric $(V,d)$ with $\depot \in V$, a \emph{profit}\footnote{We use the word ``profit'' for $\Orient$ to avoid confusion with the ``reward'' in $\StochKTSP$.} $R_v>0$ for each $v \in V$, and a budget $B>0$, the goal is to find a tour originating at $\depot$ of length at most $B$  that maximizes the collected profit.

The state-of-the-art for this NP-hard \Orient problem is a $(2+\epsilon)$-approximation algorithm~\cite{CKP12}. We denote this algorithm  as $\ALGO$ and denote the profit of the optimal \Orient tour as $\OPTO$. 
For our purposes, however, we also need to find profit at least $\OPTO$ minus an arbitrarily small additive error.
To achieve this, the tour found by our algorithm has length $O(1) \cdot B$.

\begin{restatable}{lemma}{ALGOrient} \textrm{(Bi-criteria \Orient)}
\label{lem:ALGOrient}
There is an efficient algorithm $\ALGori$ that finds a tour of length $O(1) \cdot B$ while collecting  at least $(\OPTO - \epsilon)$ profit, where $\epsilon = 1/\poly(n)$ can be made arbitrarily small.
\end{restatable}



\section{Our Approach via Critical Scaling and Repetitions}
\label{sec:OurApproach}
\subsection{A Meta-Algorithm and Critical Scaling}
\label{subsec:MetaAlg}

Our non-adaptive
$O(1)$-approximation algorithms for both $\StochKTSP$ and $\StochKCost$  have the same structure described in Meta-Algorithm~$\ALGmeta$ (Algorithm~\ref{alg:MetaAlg}). This algorithm operates in phases where it gets a budget of $O(1) \cdot \gamma^i$  in phase~$i \geq 0$ for some constant $\gamma \in (1,2)$.



In each phase~$i$, 
$\ALGmeta$ explores multiple different ``{scales}'' in the remaining graph after excluding the set $\Pi$ of vertices found in the previous phases. 
Recall, each scale corresponds to truncating the random variables at a different threshold.
This is crucial because our non-adaptive algorithm  doesn't know  the remaining reward to reach  the target $k$.
For each different scale, $\ALGmeta$ obtains a tour of length $O(1) \cdot \gamma^i$ via a sub-procedure $\ALGaux$ to which it sends  the truncated random variables as arguments
(discussed in Section~\ref{subsec:RepeatOrient}).
Eventually, $\ALGmeta$ identifies a ``critical'' scale $\jcrit$ and
appends at the end of $\Pi$ the two tours corresponding to  the scales $\jcrit$ and $\jcrit-1$. Since we only append two tours,  $\ALGmeta$ uses budget at most $O(1) \cdot \gamma^i$ in phase~$i$. 

\begin{algorithm2e}[H]
\caption{A Meta-Algorithm $\ALGmeta$}
\label{alg:MetaAlg}

 \textbf{Pre-processing stage:}\\
 set $\gamma \in (1,2), \Pi \leftarrow \emptyset$ and $\ell \leftarrow \polylog(k,n)$\;
\For{phase~$i=0,1,\cdots$}{
     set $\Pi_{i,-1} \leftarrow \emptyset$\;
    \For{scale $j=0,\cdots, \ell$}{
        set  $X_v^j \in [0,1]$ to be the truncation of $X_v$ at scale $j$ for $v \in V \setminus \Pi$, and zero  for $v \in  \Pi$ \; 
         find tour $\Pi_{i,j} \leftarrow \ALGaux \big(\big\{ X^j_v \big\}_{v \in V \setminus \Pi}~,~ i \big)$ of length $O(1) \cdot \gamma^i$\;
   }
     identify a ``critical'' scale $\jcrit$ and set $\Pi_i \leftarrow \Pi_{i,\jcrit} \cup \Pi_{i,\jcrit-1}$\;
     append tour $\Pi_i$ to $\Pi$, i.e., $\Pi \leftarrow \Pi \circ \Pi_i$\;
}
~\\
 \textbf{Probing stage:}\\
\For{phase~$i=0,1,\cdots$}{
     visit vertices in the order of $\Pi_i$ and apply certain Selection and Stopping Criteria\;
}
\textbf{Return} set of vertices selected
\end{algorithm2e}


To analyze the algorithm, we need some notation for any  phases~$i,i' \geq 1$: 
\begin{itemize}[topsep=0cm,itemsep=0cm]
    \item $\sigma_{i-1}$:  outcome of  vertices visited by $\ALGmeta$'s in the first $i-1$ phases of the probing stage.
    \item $u_{i'}(\sigma_{i-1})$:  probability that $\ALGmeta$ enters phase ${i'}+1$ in the probing stage, conditioning on $\sigma_{i-1}$.
    \item    $u_{i'}^*(\sigma_{i-1})$: probability that the cost of \OPT is more than $\gamma^{i'}$, conditioning on $\sigma_{i-1}$.

\end{itemize}
Notice $u_{i-1}(\sigma_{i-1})$ denotes the indicator variable that $\ALGmeta$ enters phase~$i$ in the probing stage.
The following Lemma~\ref{lem:KeyLemma} is the key to our theorems. 
Roughly, it says that $\ALGmeta$ is a constant approximation algorithm if it can ensure that whenever $u_i^*(\sigma_{i-1})$ is small (i.e., $\OPT$ has a large success probability within budget $\gamma^i$) then $\ALGmeta$ also succeeds with a constant probability in the first $i$ phases (i.e., only using $O(1) \cdot \gamma^i$  budget). 
The proof of Lemma~\ref{lem:KeyLemma} is standard (e.g., \cite{ENS17}),  and we defer it to Appendix~\ref{sec:MissPfOurApproach}.

\begin{restatable}{lemma}{KeyLemma} \textrm{(Key Lemma)}
\label{lem:KeyLemma}
If for some universal  constants $C > 0$,  $\gamma>1$,  any phase~$i \geq 1$, and any possible $\sigma_{i-1}$,
the algorithm  $\ALGmeta$ satisfies  
\[
u_i(\sigma_{i-1}) \leq C \cdot u_i^*(\sigma_{i-1}) + \frac{u_{i-1}(\sigma_{i-1})}{\gamma^2},
\]
 then $\ALGmeta$ is a non-adaptive $O(1)$-approximation algorithm.
\end{restatable}

All our effort will go in designing $\ALGmeta$ that satisfies the precondition of Lemma~\ref{lem:KeyLemma}.

\subsection{$\ALGaux$: Constant Repetitions of $\ALGori$ Suffice}
\label{subsec:RepeatOrient}

For a fixed scale~$j$ in phase~$i$, $\ALGmeta$ uses $\ALGaux$ (Algorithm~\ref{alg:RepeatALGori}) as a key sub-procedure to find a tour of length $O(1)\cdot\gamma^i$.
To achieve this, $\ALGaux$  runs a constant number of repetitions of $\ALGori$ on an \Orient instance where each vertex $v$ has a profit $w_v = \E[X_v]$ for input random variable $X_v \in [0,1]$ (recall, $X_v$ is the truncated random variable at scale $j$ for vertices outside $\Pi$). 
In each repetition, $\ALGaux$ excludes vertices found in previous repetitions.

\IGNORE{
The idea of repeating an \Orient algorithm appeared previously in~\cite{ENS17} where $O(\log k)$ repetitions are needed for their result.
This is the reason why their algorithms have approximation factors of $\Omega(\log k)$. 
The novelty of our approach lies in bringing the number of repetitions down to $O(1)$ by using the bi-criteria algorithm $\ALGori$ together with the scaling idea from Section~\ref{subsec:MetaAlg}.
This allows us to find a tour of length $O(1) \cdot \gamma^i$ for each scale in phase~$i$ and justifies the statement in $\ALGmeta$. \sahil{Move this discussion somewhere else?}}

\begin{algorithm2e}[H]
\caption{$\ALGaux\big(\{X_v\}_{v \in V}~,~i \big)$}
\label{alg:RepeatALGori}
\textbf{Input:}  random variables $X_v \in [0,1]$ corresponding to vertex profits and phase $i$\;
 \textbf{Main stage:}\\
 set $\gamma \in (1,2)$, $\epsilon \leftarrow 1/10^5$,  $C \leftarrow O(1)$, and $w_v=\E[X_v]$\;
 set $\Pi_i \leftarrow \emptyset$\;
\For {repetition $s = 1,\cdots, C$}{
     use $\ALGori$ to find a tour $\pi_s$ with budget $\gamma^i$, profit $\{w_v\}_{v \in V}$, and error $\epsilon$\;
     append tour $\pi_s$ to $\Pi_i$, i.e., $\Pi_i \leftarrow \Pi_i \circ \pi_s$\;
    reset $w_v = 0$ for $v \in \Pi_i$\;
   }
\textbf{Return} $\Pi_i$
\end{algorithm2e}

\medskip
\textbf{Intuition.} In the following, we prove two important properties of $\ALGaux$.
Recall from Algorithm~\ref{alg:RepeatALGori} that $\Pi_i$ denotes the union of the $C$ repetitions of $\ALGori$.
The first property (Lemma~\ref{lem:SmallT*General}) roughly says that if an \Orient tour of budget $\gamma^i$ cannot obtain much profit outside $\Pi_i$, then $\OPT$ also cannot obtain much reward outside $\Pi_i$ within budget $\gamma^i$. 
The second property (Lemma~\ref{lem:LargeT*}) roughly says that if on the other hand lots of profit can be found by an \Orient tour outside $\Pi_i$,  
then the tour $\Pi_i$ obtains a large amount of {\em expected} reward in its $C$ repetitions, much more than what $\OPT$ obtains within budget $\gamma^i$. 
This follows from the property that $\ALGori$ obtains profit close to the optimal \Orient tour.

To formally state the above two properties, we need some notation. Consider the \Orient instance in the remaining graph $V \setminus \Pi_i$ where the budget is $\gamma^i$ and each vertex $v \in V \setminus \Pi_i$ has profit $\E[X_v]$. 
Denote $\pi \subseteq V \setminus \Pi_i$ the optimal \Orient tour for this instance and  let
\begin{align}
\label{eqn:OptOrientProfit}
T := \sum_{v \in \pi} \E[X_v]
\end{align} 
be the \Orient profit obtained by $\pi$.  
For any adaptive strategy $\ADAP$, let $\Pi_i(\ADAP) \subseteq \Pi_i$ denote the random set of vertices visited by $\ADAP$ inside the tour $\Pi_i$ and  let $\overline{\Pi}_i(\ADAP) \subseteq V \setminus \Pi_i$ denote the random set of vertices visited by \ADAP outside the tour $\Pi_i$.

\begin{lemma}
\label{lem:SmallT*General}
Suppose we are given independent random variables $X_v \in [0,1]$. 
Let $T$ be as defined in~(\ref{eqn:OptOrientProfit}).
Then for any adaptive strategy $\ADAP$ which uses  at most $\gamma^i$ budget and any constant $\alpha > 1$, we have 
\[
\p\Big[ \sum_{v \in \overline{\Pi}_i(\ADAP)} X_v \geq \alpha T \Big] ~\leq~ 2 \cdot \exp\left( - \frac{(\alpha-1)^2 T/2}{1 + (\alpha-1) /3} \right).
\]
\end{lemma}

\begin{proofof}{Lemma~\ref{lem:SmallT*General}}
We construct a martingale for the (random) set $\overline{\Pi}_i(\ADAP)$ of vertices visited by $\ADAP$ in $V \setminus \Pi_i$ as follows:
When $\ADAP$ visits a vertex $v \in \overline{\Pi}_i(\ADAP)$, the martingale proceeds for one step with martingale difference defined by
\[
Z_v ~:=~ X_v - \E[X_v] ~\in~ [-1,1],
\]
and the martingale doesn't move when $\ADAP$ visits a vertex $v \in \Pi_i$.
The stopping time $\tau$ is naturally defined as the martingale step when $\ADAP$ finishes.
Since each $X_v \in [0,1]$, {the quadratic variance of the above martingale $W_{\tau} =\sum_{j=1}^\tau \E[X_j^2|\mathcal{F}_{j-1}]$ 
 is bounded by its expectation $\sum_{j=1}^\tau \E[X_j|\mathcal{F}_{j-1}]$  which is at most $T$, i.e., $\sum_{v \in \overline{\Pi}_i(\ADAP)} \E[X_v] \leq T$.}
Therefore, applying Freedman's inequality (Theorem~\ref{thm:FreedmanInequality}),  we have
\[
\p\Big[ \sum_{v \in \overline{\Pi}_i(\ADAP)} X_v \geq \alpha T \Big] 
~~\leq~~ \p\Big[\big| \sum_{v \in \overline{\Pi}_i(\ADAP)} Z_v \big| \geq (\alpha - 1)T \wedge W_{\tau} \leq T \Big] ~~\leq~~ 2 \cdot \exp\left( - \frac{(\alpha-1)^2 T/2}{1 + (\alpha-1) /3} \right).
\]
This finishes the proof of Lemma~\ref{lem:SmallT*General}.
\end{proofof}

\begin{lemma}
\label{lem:LargeT*}
Suppose we are given independent random variables $X_v \in [0,1]$. 
Let $T$ be as defined in~(\ref{eqn:OptOrientProfit}). 
Then for any adaptive strategy $\ADAP$ which uses budget at most $\gamma^i$, we have
\[
\sum_{v \in \Pi_i \setminus \Pi_i(\ADAP)} \E[X_v] ~\geq~ (C-1) (T - \epsilon) - \epsilon.
\]
\end{lemma}

\begin{proofof}{Lemma~\ref{lem:LargeT*}}
Since $\ADAP$ uses budget at most $\gamma^i$, the set $\Pi_i(\ADAP) \subseteq \Pi_i$ can always be visited by a tour of length at most $\gamma^i$. 
{Consider the first tour $\pi_1$ found by Algorithm~\ref{alg:RepeatALGori}. 
Since the tour with length at most $\gamma^i$ that visits the set $\Pi_i(\ADAP)$  is a valid \Orient tour when $\pi_1$ is found,
it follows from Lemma~\ref{lem:ALGOrient} that $\sum_{v \in  \Pi_i(\ADAP)} \E[X_v]  \leq \epsilon + \sum_{v \in \pi_1} \E[X_v] $.}
For any $s \in  \{2,3,\cdots,C\}$, since $\pi$ is an \Orient tour in $V \backslash \Pi_i$ of length at most $\gamma^i$ with $\sum_{v \in  \pi} \E[X_v] = T$, Lemma~\ref{lem:ALGOrient} implies that $\sum_{v \in \pi_s} \E[X_v] \geq T - \epsilon$. 
Therefore, a simple calculation gives
\[
\sum_{v \in \Pi_i \setminus \Pi_i(\ADAP)} \E[X_v] \quad = \quad  \sum_{v \in \Pi_i} \E[X_v] - \sum_{v \in \Pi_i(\ADAP)} \E[X_v] \quad  \geq \quad  (C-1)(T - \epsilon) - \epsilon,
\]
which finishes the proof of Lemma~\ref{lem:LargeT*}.
\end{proofof}

\IGNORE{
For $s \in [C]$, define $\Pi_s := \bigcup_{t=1}^s \pi_t$ the tours found in the first $s$ repetitions. 
Consider the first repetition $s_0 \in [C]$ such that 
$\sum_{v \in \Pi_i(\ADAP) \setminus \Pi_s} \E[X_v] \leq T/2$.
If such $s_0$ doesn't exist, we will define $s_0$ to be $C$.
For any $s <  s_0$, since $\Pi_i(\ADAP) \setminus \Pi_{s-1}$ can be visited by a tour of length at most $\gamma^i$, it follows from Lemma~\ref{lem:ALGOrient} that \[
\sum_{v \in \pi_s} \E[X_v] ~\geq~ \Big(\sum_{v \in \Pi_i(\ADAP) \setminus \Pi_{s-1}} \E[X_v] \Big) - \epsilon.
\]
Since $\pi_s \cap \Pi_{s-1} = \emptyset$, this implies that 
\[
\sum_{v \in \pi_s \setminus \Pi_i(\ADAP)} \E[X_v] ~\geq~ \Big( \sum_{v \in \Pi_i(\ADAP) \setminus \Pi_s} \E[X_v]  \Big) - \epsilon ~\geq~ T/2 - \epsilon.
\] 

For any $s_0 < s \leq C$, since $\pi^*$ is a valid \Orient tour with budget $\gamma^i$ and value $T$, we have from Lemma~\ref{lem:ALGOrient} that $\sum_{v \in \pi_s} \E[X_v] \geq T - \epsilon$.
So it follows that 
\[
\sum_{v \in \pi_s \setminus \Pi_i(\ADAP)} \E[X_v] ~\geq~ \sum_{v \in \pi_s} \E[X_v] - \sum_{v \in \Pi_i(\ADAP) \setminus \Pi_{s-1}} \E[X_v] ~\geq~ T/2 - \epsilon.
\]
Therefore, we conlcude that whenever $s \neq s_0$, we have $\sum_{v \in \pi_s \setminus \Pi_i(\ADAP)} \E[X_v] \geq T/2 - \epsilon$. 
It follows that 
\[
\sum_{v \in \Pi \setminus \Pi_i(\ADAP)} \E[X_v] ~\geq~ \sum_{s \neq s_0} \sum_{v \in \pi_s \setminus \Pi_i(\ADAP)} \E[X_v] ~\geq~ (C-1)(T/2 - \epsilon).
\]
}


\section{\StochKTSP}
\label{sec:StochKTSP}
In this section we prove Theorem~\ref{thm:StochKTSP}, which is restated below for convenience.

\stochkTSP*

To prove this theorem we  carefully choose the parameters of our Meta-Algorithm from last section.

\subsection{The Algorithm}

We assume without loss of generality that the stochastic reward $R_v \leq k$ almost surely for each vertex $v \in V$.
Our algorithm \ALGStochKTSP for \StochKTSP problem is given in Algorithm~\ref{alg:StochKTSP}.
It is an instantiation of the Meta-Algorithm \ALGmeta (Algorithm~\ref{alg:MetaAlg}) by setting the phase parameter $\gamma = 1.1$, number of scales $\ell = \lfloor \log k \rfloor$, and number of repetitions $C = 6000$.
For each scale $j$ in phase $i$, we set the random variable $X_v^j$ for any vertex $v \in V \setminus \Pi$ in \ALGmeta to be the stochastic reward $R_v$ truncated at $k/2^j$ and then scaled down to $[0,1]$. 
We identify the ``critical'' scale $\jcrit$ in \ALGmeta as follows:
For each scale $j$ in phase $i$, denote $\Pi_{i,j}$ the $C=6000$ repetitions of $\ALGori$ and let $T_{i,j}$ be the profit obtained by the $3$-approximation \Orient algorithm \ALGO in $V \setminus (\Pi \cup \Pi_{i,j})$. 
The ``critical'' scale $\jcrit$ is the smallest scale $j$ such that $T_{i,j} \geq 1/300$, i.e., sufficient profit remains outside even after $C$ repetitions. 
We add the two tours corresponding to the ``rich'' scale $\jcrit$ and the  ``poor'' scale $\jcrit-1$ into $\Pi$.
In the case when there is no  scale $j$ with $T_{i,j} \geq 1/300$, we simply set $\jcrit$ to be the last scale $\ell$. 
In the probing stage, the selection and the stopping criteria are straightforward: we collect reward from every  visited vertex  and stop when the total reward reaches $k$. 


\begin{algorithm2e}[H]
\caption{\ALGStochKTSP for \StochKTSP problem}
\label{alg:StochKTSP}

 \textbf{Pre-processing stage:}\\
 set $\gamma \leftarrow 1.1, \epsilon \leftarrow 1/10^5, \Pi \leftarrow \emptyset$, $\ell=\lfloor \log k\rfloor$, and  $C \leftarrow 6000$  \;
\For{phase $i=0,1,\cdots$}{
     set $\Pi_{i,-1} \leftarrow \emptyset$ \;
    \For{scale $j=0,\cdots, \ell$}{
          set profit   $w^j_v= \E\left[\min\left\{R_v \cdot 2^j/k, 1\right\} \right] \cdot \one[v\in V\setminus \Pi]$ ; {\color{gray} \qquad /* Scale $j$ truncates at $k/2^j$ */} \\
         set $\Pi_{i,j} \leftarrow \emptyset$ \;
        \For{repetition $s =1,2,\cdots,C$           {\color{gray} \qquad /* Constant repetitions of $\ALGori$ */}\\}{
             use $\ALGori$ to find tour $\pi_{i,j,s}$ with budget $\gamma^i$, profit $\{w_v^j\}_{v \in V}$, and error $\epsilon$ \;
             append tour $\pi_{i,j,s}$ to $\Pi_{i,j}$, i.e., $\Pi_{i,j} \leftarrow \Pi_{i,j} \circ \pi_{i,j,s}$ \;
             reset $w^j_v =0$ for $v \in \Pi_{i,j}$  \;
        }
          ~{\color{gray} /* Check whether $j$ is a ``critical'' scale */}\\
         use $\ALGO$ to find tour $\pi_{i,j}^{\ori}$ with budget $\gamma^i$ and profit $\{w_v^j\}_{v \in V}$ \;
         set $T_{i,j} \leftarrow \sum_{v\in\pi_{i,j}^{\ori}} w^j_v$  \;
         
        \If {$T_{i,j}\geq 1/300$ or $j= \ell$   }{
         append tour $\Pi_i := \Pi_{i,j} \cup \Pi_{i,j-1}$ to $\Pi$, i.e., $\Pi \leftarrow \Pi \circ \Pi_i$  \; 
         \textbf{Break} \;
         }
	}
}
~\\
 \textbf{Probing stage:}\\
\For{phase $i=0,1,\cdots$}{
     visit vertices in the order of $\Pi_i$ and apply the following selection and stopping criteria \;
     \textbf{Selection Criterion:} select every vertex visited  \;
     \textbf{Stopping Criterion:} total reward reaches $k$ \;
}
\textbf{Return} set of vertices selected
\end{algorithm2e}

\subsection{Proof of Theorem~\ref{thm:StochKTSP}}

Recall from Section~\ref{subsec:MetaAlg} that to prove Theorem~\ref{thm:StochKTSP}, we only need to prove the precondition in Lemma~\ref{lem:KeyLemma}. 
The remainder of this section proves this precondition  for \ALGStochKTSP as given in the following Lemma~\ref{lem:KeyLemStochKTSP}.
\begin{lemma}
\label{lem:KeyLemStochKTSP}
For $\gamma = 1.1$, any phase $i > 0$ in the probing stage of \ALGStochKTSP  satisfies
\begin{align}
\label{eqn:KeyLemStochKTSP}
u_i(\sigma_{i-1}) \leq 100 \cdot u_i^*(\sigma_{i-1}) + \frac{u_{i-1}(\sigma_{i-1})}{\gamma^2}.
\end{align}
\end{lemma}

Before proving Lemma~\ref{lem:KeyLemStochKTSP}, we discuss the high-level intuition of our proof. 

\noindent \textbf{Intuition.} 
Assume without loss of generality that $u_i^*(\sigma_{i-1}) < 0.01$, i.e., $\OPT$ finds $k$ reward within budget $\gamma^i$ with probability at least $0.99$, as otherwise the lemma trivially holds. 
Now the plan is to show that with constant probability, $\ALGStochKTSP$ finds more reward than $\OPT$ restricted to budget $\gamma^i$, even when all rewards in $\sigma_{i-1}$ are given to $\OPT$ for free.
This allows us to focus on the remaining graph with vertex set $V_i := V \setminus \sigma_{i-1}$ where we repeat $\ALGori$ for different scales.

Our arguments rely on the notion of ``richness''. 
We call a scale $j$ ``rich'' if $T_{i,j} \geq 1/300$ and otherwise ``poor''. 
A scale being poor indicates that not much reward can be collected outside the $C$ repetitions of $\ALGori$ for that scale, in which case we can use Lemma~\ref{lem:SmallT*General} to argue that $\OPT$ cannot find much reward outside. 
A scale being rich implies that each repetition of $\ALGori$ for  that scale finds a significant amount of reward, in which case we can apply Lemma~\ref{lem:LargeT*} to argue that $\ALGStochKTSP$ finds much more reward than $\OPT$ in the $C$ repetitions for that scale. 
Our critical scale $\jcrit$ corresponds to the transition from poor to rich scales. Since the algorithm includes both $\jcrit$ and $\jcrit-1$, roughly the reason why our analysis works is that  we use the poor scale $\jcrit -1$ to argue $\OPT$ cannot find much reward outside  our tours and we use the rich scale $\jcrit$ to argue $\OPT$ cannot find much more reward inside. The 
final analysis has to do some case analysis depending on whether the transition ever happens or not.

\begin{proofof}{Lemma~\ref{lem:KeyLemStochKTSP}}
We fix any outcome $\sigma_{i-1}$ of vertices visited by $\ALGStochKTSP$ in the first $i-1$ phases in its probing stage. 
The lemma trivially holds in the case where $u_i^*(\sigma_{i-1}) \geq 0.01$ as we have $100u_i^*(\sigma_{i-1}) \geq 1$. 
If $u_{i-1}(\sigma_{i-1}) = 0$ which means that \ALGStochKTSP already collects reward $k$ before entering phase $i$ in the probing stage, then $u_i(\sigma_{i-1})=0$ and again the lemma trivially holds.
We therefore assume that $u_i^*(\sigma_{i-1}) < 0.01$ and that $u_{i-1}(\sigma_{i-1})=1$.
Now proving Lemma~\ref{lem:KeyLemStochKTSP} is equivalent to proving
\begin{align}
\label{eqn:ALGStochKTSPAssume}
u_i(\sigma_{i-1}) \leq 100 u_i^*(\sigma_{i-1}) + 1/\gamma^2.
\end{align}
To prove~(\ref{eqn:ALGStochKTSPAssume}), we need the following notation.
Denote $V_i \coloneqq V \setminus \sigma_{i-1}$ the vertex set of the remaining graph where vertices in $\sigma_{i-1}$ are excluded.
Denote $\overline{\Pi}_i(\OPT) \subseteq V_i \setminus \Pi_i$ the (random) set of vertices visited by $\OPT$ outside $\sigma_{i-1} \cup \Pi_i$ within budget $\gamma^i$, and denote $\Pi_i(\OPT) \subseteq \Pi_i$ the (random) set of vertices visited by $\OPT$ inside $\Pi_i$.
We consider three cases:

\noindent \textbf{Case (1): (Scale $0$ is rich)} $T_{i,0} \geq 1/300$.
In this case, our algorithm appends tour $\Pi_i := \Pi_{i,0}$ to $\Pi$ (recall that $\Pi_{i,-1} = \emptyset$), and these will be the phase $i$ vertices visited in the probing stage. 
We show that  $T_{i,0} \geq 1/300$ implies that each repetition of $\ALGori$ has large expected reward (notice the random rewards are not truncated at scale $0$). As we repeat $\ALGori$ for $C = 6000$ times, the tour $\Pi_{i,0}$ has expected reward much larger than the target $k$. 

Since $T_{i,0}$ is the profit of a valid \Orient tour with length at most $\gamma^i$, for each $s \in \{1,\ldots,C\}$ we have $T_{i,0,s} \geq T_{i,0} - \epsilon \geq 1/300 - \epsilon$, where $\epsilon = 1/10^5$ is the small error term for \ALGori in Lemma~\ref{lem:ALGOrient}.
Thus, 
\[
\sum_{v \in \Pi_{i,0}} \E\left[\min\left\{R_v/k, 1\right\}\right] \quad \geq \quad 20 - 6000 \epsilon \quad \geq \quad 19.
\]
Notice that $R_v/k \in [0,1]$, so applying Chernoff bound (Theorem~\ref{thm:ChernoffBound}) we have
\[
1-u_i(\sigma_{i-1}) \quad \geq \quad \p \Big[\sum_{v \in \Pi_{i,0}} R_v \geq k \Big] \quad = \quad \p\Big[ \sum_{v \in \Pi_{i,0}} \min\left\{R_v/k, 1\right\} \quad \geq \quad 1 \Big] \quad \geq \quad 0.9,
\]
which means $u_i(\sigma_{i-1}) \leq 0.1 \leq 1/\gamma^2$. 
This proves~(\ref{eqn:ALGStochKTSPAssume}) and finishes the proof of Lemma~\ref{lem:KeyLemStochKTSP} in this case.

\noindent \textbf{Case (2): (Scale $\ell$ is poor)} $T_{i,j} \leq 1/300$ for every scale $j = 0,\cdots,\ell$.  
In this case, our algorithm adds $\Pi_i := \Pi_{i,\ell} \cup \Pi_{i,\ell-1}$ to $\Pi$, and these will be the phase $i$ vertices visited in the probing stage. 
We argue that the assumption of $T_{i,\ell} \leq 1/300$ implies that with constant probability  $\OPT$ finds {\em no} reward outside $\sigma_{i-1} \cup \Pi_i$ within budget $\gamma^i$.
If $\OPT$ still manages to find $k$ reward, then all $k$ reward must come from vertices in $\sigma_{i-1} \cup \Pi_i$, in which case $\ALGStochKTSP$ also finds $k$ reward.
  In this case our argument already works with the vertices $\Pi_{i,\ell}$ added to $\Pi$, i.e., we do not even need vertices in $\Pi_{i, \ell-1}$.

Since $\ALGO$ is a 3-approximation \Orient algorithm, for any set of vertices $S \subseteq V_i \setminus \Pi_i$ that can be visited within budget $\gamma^i$, we have $\sum_{v \in S} \E \left[ \min \left\{ R_v \cdot 2^\ell/k, 1 \right \} \right] \leq 3T_{i,\ell} \leq  0.01$.
Since $\ell = \lfloor \log k\rfloor$, we have $k/2^\ell \in [1,2]$, and therefore 
\begin{align}
\label{eqn:YiSumSmall}
\sum_{v \in S} \E[\min\{R_v, 1\}] 
\quad \leq \quad \sum_{v \in S} \E\left[ \min \left \{ 2\cdot R_v \cdot 2^\ell/k, 1 \right\} \right]  \quad \leq \quad 0.02. 
\end{align}
Recall that $\overline{\Pi}_i(\OPT) \subseteq V_i \setminus \Pi_i$ denotes the (random) set of vertices visited by $\OPT$ outside $\sigma_{i-1} \cup \Pi_i$ within budget $\gamma^i$.
Since each $\min\{R_v, 1\} \in \{0,1\}$, 
the best probability of obtaining truncated reward at least 1 in $V_i \setminus \Pi_i$ within budget $\gamma^i$ is achieved by a non-adaptive strategy.
Therefore, Markov's inequality together with~(\ref{eqn:YiSumSmall}) implies that
\[
\p\Big[ \sum_{v \in \overline{\Pi}_i(\OPT)} \min\{R_v, 1\} \geq 1 \Big] ~\leq ~ 0.02.
\]

Since $u_i^*(\sigma_{i-1}) < 0.01$, we have that with probability at least $1 - 0.01 - 0.02 = 0.97$, $\OPT$ finds $k$ reward in $V$ but $0$ reward outside $\sigma_{i-1} \cup \Pi_i$.
In this case, \ALGStochKTSP also finds $k$ reward among vertices visited in the first $i$ phases.
So we have $u_i(\sigma_{i-1}) \leq 1-0.97=0.03 \leq 1/\gamma^2$, which establishes~(\ref{eqn:ALGStochKTSPAssume})  in this case.

\noindent \textbf{Case (3): (Transition from poor to rich scale at $\jcrit$)} $T_{i,0} \leq 1/300$ but $T_{i,j} > 1/300$ for some $j \in [\ell]$. In this case, let $\jcrit = \min \big\{j \in [\ell]: T_{i,j} > 1/300 \big\}$ be our critical scale. The algorithm appends $\Pi_i := \Pi_{i,\jcrit-1} \cup \Pi_{i,\jcrit}$ to $\Pi$.
To prove that \ALGStochKTSP will not continue to phase $i+1$ with constant probability, we show that the following two events happen with constant probability: 
\begin{enumerate}[topsep=0cm,itemsep=0cm]
\item  $\OPT$ doesn't find too much reward outside $\sigma_{i-1} \cup \Pi_i$ within budget $\gamma^i$.
\item  $\ALGStochKTSP$ finds much more reward inside $\Pi_i$ than $\OPT$ does since it is restricted to budget $\gamma^i$. 
\end{enumerate}
We show that the first event follows from $T_{i,\jcrit-1} \leq 1/300$ while the second event follows from $T_{i,\jcrit} \geq 1/300$.
From these we conclude that with constant probability, $\ALGStochKTSP$ obtains at least as much reward as $\OPT$ restricted to budget $\gamma^i$.

We first argue that $\OPT$ doesn't find too much reward outside $\sigma_{i-1} \cup \Pi_i$. Specifically, we prove that 
\begin{align}
\label{eqn:OPTPoorOutside}
\p\Big[ \sum_{v \in \overline{\Pi}_i(\OPT)} R_v < k/2^{\jcrit-1} \Big] ~\geq~ 0.5.
\end{align}
Notice that $T_{i,\jcrit-1} \leq 1/300$ together with the fact that $\ALGO$ is a $3$-approximation for \Orient implies that for any set of vertices $S \subseteq V_i \setminus \Pi_i$ that can be visited within distance $\gamma^i$, we have
\[
\sum_{v \in S} \E\left[\min \left\{R_v \cdot 2^{\jcrit-1}/k, 1 \right\} \right] \leq 0.01.
\]
Since the random variables $\min \big\{R_v \cdot 2^{\jcrit-1}/k, 1 \big\} \in [0,1]$, applying Lemma~\ref{lem:SmallT*General} we have
\[
\p\Big[ \sum_{v \in \overline{\Pi}_i(\OPT)} \min \big\{R_v \cdot 2^{\jcrit-1}/k, 1 \big\} \geq 1 \Big] \quad \leq \quad 2\exp\left( -\frac{0.99^2/2}{0.01+0.99/3} \right) \quad \leq \quad 0.5,
\]
which immediately implies~(\ref{eqn:OPTPoorOutside}).

Now we  argue that inside $\Pi_i$, we find much more reward than $\OPT$ does when it's restricted to budget $\gamma^i$. 
Specifically, we prove  that
\begin{align}
\label{eqn:RicherThanOPTInside}
\p\Big[ \sum_{v \in \Pi_i \setminus \Pi_i(\OPT)} R_v \geq k/2^{\jcrit-1} \Big] ~\geq~ 0.8,
\end{align}
where recall that $\Pi_i(\OPT) \subseteq \Pi_i$ is the (random) set of vertices visited by $\OPT$ inside $\Pi_i$ within budget $\gamma^i$.
Notice that $T_{i,\jcrit} > 1/300$ together with Lemma~\ref{lem:LargeT*} implies that 
\[
\sum_{v \in \Pi_i \setminus \Pi_i(\OPT)} \E[\min\{R_v \cdot 2^\jcrit/k, 1\}] \quad \geq \quad (6000 - 1) \cdot (T_{i,\jcrit} - \epsilon) - \epsilon \quad \geq \quad 19.
\]
Applying Chernoff bound (Theorem~\ref{thm:ChernoffBound}), we have
\[
\p\Big[ \sum_{v \in \Pi_i \setminus \Pi_i(\OPT)} \min\{R_v \cdot 2^\jcrit/k, 1\} \geq 2 \Big] ~\geq~ 0.8,
\]
 which implies~(\ref{eqn:RicherThanOPTInside}).

Now we complete the proof of~(\ref{eqn:ALGStochKTSPAssume}) in this final case. 
From~(\ref{eqn:OPTPoorOutside}) and~(\ref{eqn:RicherThanOPTInside}) and our assumption that $u_i^*(\sigma_{i-1}) < 0.01$, we have that with probability at least $1 - 0.5 - 0.2 - 0.01 \geq 1/4$, 
all the following three events hold: (1)~$\sum_{v \in \overline{\Pi}_i(\OPT)} R_v < k/2^{\jcrit-1}$, (2)~$\sum_{v \in \Pi_i \setminus \Pi_i(\OPT)} R_v \geq k/2^{\jcrit-1}$, and (3)~$\OPT$ obtains at least $k$ reward within budget $\gamma^i$.
When all these three events hold, $\ALGStochKTSP$ also finds at least $k$ reward before visiting any vertex from phase $i+1$. 
Therefore, $u_i(\sigma_{i-1}) \leq 3/4 \leq 1/\gamma^2$, and this completes the proof of Lemma~\ref{lem:KeyLemStochKTSP}.
\end{proofof}

\section{\StochKCost}
\label{sec:StochKCost}

In this section we prove Theorem~\ref{thm:StochKCost}, which is restated below for convenience. Throughout this section, we will remove the restriction that a vertex $v$ can only be selected  if we are currently at  $v$ because this is equivalent to the original problem up to a factor of $2$.

\stochkCost*

\IGNORE{\haotian{Don't restate the problem, use one sentence to explain.}
As noted in Footnote~\ref{footnote:DistSelect}, up to a factor of $2$, one can remove the restriction that a vertex can only be selected if we are currently at that vertex. 
Formally, we work with this equivalent (up to a constant factor) problem which we also call \StochKCost: 
\noindent \textbf{\StochKCost:}  In the \StochKCost problem, we are given a metric $(V, d)$ with a $\depot \in V$. Each vertex $v \in V$ has an independent stochastic cost $C_v \in \mathbb{R}_{\geq 0}$. All cost distributions are given as input but the actual cost instantiation $C_v$ is only known when vertex $v$ is visited and stay fixed afterwards.
A vertex $v$ can be selected {\em if and only if} $v$ is visited.
The goal is to  adaptively find a tour $\Pi$ originating from $\depot$ that selects a set $S$ of $k$ visited vertices while minimizing the expected {\em total} cost
\[
\E \Big[d(\Pi) + \sum_{v \in S} C_v \Big],
\]
where the first term $d(\Pi)$ is the total length of the tour $\Pi$ and the second term is the total cost of the items selected.}

Recall, an additional challenge for  \StochKCost  is that there is no obvious way to truncate the cost distributions $C_v$, even if the remaining target $k'$ is known.
Truncating at the ``average'' cost per remaining reward (i.e., $\p[C_v \leq O(\gamma^i/k')]$) will fail when some vertices in the optimal tour have costs much smaller than $O(\gamma^i/k')$, while the other vertices have much higher costs. 
We overcome this by considering $\p[C_v \leq O(\gamma^i/2^j)]$ for all possible scales $j \in \{0,\cdots, i\cdot \log \gamma + \log n\}$. To identify a ``critical'' scale, we evaluate the maximum target a tour at a given scale can get with constant probability within cost budget $2^i$. 
We show this can be approximately computed  via dynamic programming.

The rest of this section is devoted to proving Theorem~\ref{thm:StochKCost}.

\subsection{The Algorithm}
Our algorithm $\ALGStochKCost$ for \StochKCost is given in Algorithm~\ref{alg:StochKCost}. 
$\ALGStochKCost$ is an instantiation of our Meta-Algorithm \ALGmeta in Algorithm~\ref{alg:MetaAlg} by setting the phase parameter $\gamma = 1.1$ and number of repetitions $C = 6000$.
The number of scales in phase $i \geq 0$ will be $\ell_i = \lfloor i \cdot \log \gamma + \log n \rfloor$. (Notice, unlike \StochKTSP, the number of scales changes with phase.)
For scale $j$ in phase $i$, we set a random variable $X_v^j$ for  vertex $v \in V \setminus \Pi$ in \ALGmeta to be the indicator variable that cost $C_v \leq  \gamma^i/2^j$.

We identify a ``critical'' scale $\tildejcrit$  as follows:
For any phase $i \geq 0$ and scale $j \in \{0,\cdots,\ell_i\}$, define $Y_{i,j} \in \mathbb{Z}_{\geq 0}$ to be the maximum number of vertices that can be selected from $\Pi_{i,j} \cup \Pi_{i,j-1}$ with probability at least $0.2$ within cost budget $3\gamma^i$.
Ideally, we want to compute $Y_{i,j}$ for every scale $j \in \{0,\cdots, \ell_i\}$ and set the critical scale to be the one that maximizes $Y_{i,j}$.
Unfortunately, $Y_{i,j}$ cannot be computed efficiently as the corresponding problem is NP-Hard. 
To get around this issue, we compute an approximate value $\tildeY_{i,j}$ in Step~\ref{algstep:DP} of \ALGStochKCost via a dynamic programming sub-procedure $\ALGDP$. 
We discuss the details of $\ALGDP$  in Section~\ref{subsec:ALGDP},
where we prove the following Lemma~\ref{lem:DPestimate} which roughly says that the $\tildeY_{i,j}$ computed by $\ALGStochKCost$ is a reasonably good approximation of $Y_{i,j}$.
\begin{restatable}{lemma}{DPestimate}
\label{lem:DPestimate}
For any phase $i \geq 0$ and any scale $j \in \{0,\cdots, \ell_i\}$, the approximate value $\tildeY_{i,j}$ computed in Step~\ref{algstep:DP} of \ALGStochKCost satisfies that (1) $\tildeY_{i,j} \geq Y_{i,j}$, and (2) $\tildeY_{i,j}$ vertices can be selected from $\Pi_{i,j} \cup \Pi_{i,j-1}$ within cost budget $6 \gamma^i$ with probability at least $0.2$.
\end{restatable}
After computing $\tildeY_{i,j}$ for each scale $j \in \{0,\cdots,\ell_i\}$, we simply set $\tildejcrit$ to be the scale that maximizes $\tildeY_{i,j}$ and add the two tours corresponding to scales $\tildejcrit-1$ and $\tildejcrit$ into $\Pi$.

In the probing stage, the Stopping Criterion is natural: we stop whenever the number of selected vertices reaches $k$. 
The selection process needs some care since not all vertices visited in the previous phases are selected. 
Our algorithm therefore runs two selection processes consecutively:
In Selection-Process~1, we select as many unselected vertices from those visited in the previous phases within total cost  $\gamma^i$.
This is to ensure that we select from $\sigma_{i-1}$ at least as many vertices as $\OPT$ restricted to budget $\gamma^i$.
In Selection-Process~2, we select as many vertices as possible from $\Pi_i$ within total cost $6\gamma^i$. 
The above Lemma~\ref{lem:DPestimate} guarantees that at least $\tildeY_{i,\tildejcrit}$ vertices can be selected in this process with probability at least $0.2$.


\begin{algorithm2e}
\caption{\ALGStochKCost for \StochKCost problem}
\label{alg:StochKCost}

 \textbf{Pre-processing stage:}\\
 set $\gamma \leftarrow 1.1$, $\epsilon \leftarrow 1/10^5$, $\Pi \leftarrow \emptyset$ and $C \leftarrow 6000$ \;
\For{phase $i=0,1,\cdots$}{
     set $\Pi_{i,-1} \leftarrow \emptyset$ \;
       \For{scale $j=0,\cdots, \ell_i$, where $\ell_i =\lfloor i \cdot \log \gamma + \log n \rfloor$}{
         set profit $w^j_v = \p[C_v\leq \gamma^i/2^j] \cdot \one[v\in V\setminus \Pi]$ ; {\color{gray} \qquad /* Scale $j$ ``truncates'' at $\gamma^i/2^j$ */}\\
         set $\Pi_{i,j} \leftarrow \emptyset$ \;
        \For{repetition $s =1,2,\cdots,C$ {\color{gray} \qquad \text{/* Constant repetitions of $\ALGori$ */} }\\}{
             use $\ALGori$ to find tour $\pi_{i,j,s}$ with budget $\gamma^i$, profit $\{w_v^j\}_{v \in V}$ and error $\epsilon$ \;
             append tour $\pi_{i,j,s}$ to $\Pi_{i,j}$, i.e., $\Pi_{i,j} \leftarrow \Pi_{i,j} \circ \pi_{i,j,s}$ \;
             reset $w^j_v=0$ for $v \in \Pi_{i,j}$ \;
        }
         ~{\color{gray} /* Approximately compute the maximum number of vertices that can be selected from $\Pi_{i,j} \cup \Pi_{i,j-1}$ within cost budget $3\gamma^i$ and with probability at least 0.2 */}\\
         find the largest integer $\tildeY_{i,j} \leq n$ such that  $\ALGDP(\tildeY_{i,j}, 3\gamma^i, \Pi_{i,j}\cup \Pi_{i,j-1}) \geq 0.2$ \; \label{algstep:DP}
    }
    ~{\color{gray} /* Identify a ``critical'' scale */}\\
     set $\tildejcrit \leftarrow \arg\max_{j} \tildeY_{i,j}$ and $\Pi_i \leftarrow \Pi_{i,\tildejcrit} \cup \Pi_{i,\tildejcrit-1}$ \;
     append tour $\Pi_i$ to $\Pi$, i.e., $\Pi \leftarrow \Pi \circ \Pi_i$ \;
}
~\\

 \textbf{Probing stage:}\\
\For{phase $i=0,1,\cdots$}{
     set $\sigma_{i-1} \leftarrow \bigcup_{t=0}^{i-1} \Pi_t$  \;
     visit vertices in the order of $\Pi_i$ and apply the following selection and stopping criteria \;
     ~{\color{gray} /* Select (unselected) vertices visited in previous phases */}\\
     \textbf{Selection-Process~1:} select as many vertices as possible from $\sigma_{i-1}$ within total cost $\gamma^i$ \; 
     ~{\color{gray} /* Select vertices visited in the current phase */}\\
     \textbf{Selection-Process~2:} select as many vertices as possible from $\Pi_i$ within total cost $6 \gamma^i$  \; 
     \textbf{Stopping Criterion:} total number of vertices selected reaches $k$ \;
}
\textbf{Return} the set of selected vertices
\end{algorithm2e}

\subsection{Proof of Theorem~\ref{thm:StochKCost}}
Recall from Section~\ref{subsec:MetaAlg} that to prove Theorem~\ref{thm:StochKCost}, we only need to prove the precondition in Lemma~\ref{lem:KeyLemma}. 
The remainder of this section proves this precondition for \ALGStochKCost as stated in the following Lemma~\ref{lem:KeyLemStochKCost}.

\begin{restatable}{lemma}{KeyLemStochKCost}
\label{lem:KeyLemStochKCost}
For  $\gamma = 1.1$, any phase $i > 0$ in the probing stage of \StochKCost satisfies
\begin{align}
\label{eqn:ALGStochKCost}
u_i(\sigma_{i-1}) \leq 100 u_i^*(\sigma_{i-1}) + \frac{{u_{i-1}(\sigma_{i-1})}}{\gamma^2}.
\end{align}
\end{restatable}

Before proving Lemma~\ref{lem:KeyLemStochKCost}, we need some notation.

\noindent \textbf{Notation.} For any (possibly adaptive) algorithm $\ADAP$, we say $\ADAP$ has a {\em distance budget} of $B$ if it is allowed to travel a total distance of at most $B$; we say $\ADAP$ has a {\em cost budget} of $B$ if it is allowed to select vertices up to a total cost of $B$;
we say $\ADAP$ has a {\em total budget} of $B$ if its total distance travelled {\em plus} the total cost of selecting vertices is restricted to be at most $B$.
An algorithm satisfying a  budget constraint is said to be {\em within} that budget.
For any phase $i$, denote $V_i := V \setminus \sigma_{i-1}$ the set of vertices in the remaining graph where vertices in $\sigma_{i-1}$ are excluded.
For any fixed outcome $\sigma_{i-1}$ and any target $Y \in \mathbb{Z}_{\geq 0}$ , denote $p^*_{i,Y}(\sigma_{i-1})$ the probability that $\OPT$ selects
at least $Y$ vertices from $V_i$ within total budget $\gamma^i$ and $p_{i,Y}(\sigma_{i-1})$ the probability that the tour $\Pi_i \subseteq V_i$ found by \ALGStochKCost contains $Y$ vertices which can be selected within cost budget $6\gamma^i$. 

The proof of Lemma~\ref{lem:KeyLemStochKCost} relies on the following Lemma~\ref{lem:KeyLemmaAux}, which says that if $\OPT$ selects $Y$ vertices in $V_i$ within total budget $\gamma^i$ with probability at least $0.9$, then we can select $Y$ vertices in $\Pi_i$ within cost budget $6\gamma^i$ with probability at least $0.2$.
The proof of Lemma~\ref{lem:KeyLemStochKCost} from Lemma~\ref{lem:KeyLemmaAux} is standard, see  Appendix~\ref{sec:MissPfStochKCost}.

\begin{restatable}{lemma}{KeyLemmaAux}
\label{lem:KeyLemmaAux}
For any phase $i \geq 0$, any target $Y \in \mathbb{Z}_{\geq 0}$ and any outcome $\sigma_{i-1}$ of vertices visited in the previous $i-1$ phases, if $p^*_{i,Y}(\sigma_{i-1}) \geq 0.9$ then we have $p_{i,Y}(\sigma_{i-1}) \geq 0.2$.
\end{restatable}

We need some notation to prove Lemma~\ref{lem:KeyLemmaAux}.

\noindent \textbf{Notation.}  We say a vertex $v \in V$ is {\em qualified} for scale $j \in \{0,\cdots, \ell_i\}$ if its cost $C_v \leq \gamma^i/2^j$.  
For each scale $j \in \{0,\cdots, \ell_i\}$, denote $\pi_{i,j}^*$ the optimal \Orient tour in $V_i \setminus \Pi_{i,j}$ with budget $\gamma^i$ 
where each vertex $v \in V_i \setminus \Pi_{i,j}$ has profit $\p[C_v \leq \gamma^i/2^j]$. Define $T_{i,j}^* := \sum_{v \in \pi_{i,j}^*} \p[C_v \leq \gamma^i/2^j]$ and $T_{i,j} := \sum_{v \in \Pi_{i,j}} \p[C_v \leq \gamma^i/2^j]$ to be the total \Orient profit of tour $\pi_{i,j}^*$ and $\Pi_{i,j}$, respectively. Denote $\overline{\Pi}_{i,j}(\OPT) \subseteq V_i \setminus (\Pi_{i,j}\cup\Pi_{i,j-1})$ the (random) set of vertices visited by $\OPT$ outside $\sigma_{i-1} \cup (\Pi_{i,j}\cup\Pi_{i,j-1})$ within total budget $\gamma^i$, and $\Pi_{i,j}(\OPT) \subseteq \Pi_{i,j}\cup\Pi_{i,j-1}$ the (random) set of vertices visited by $\OPT$ inside $\Pi_{i,j}\cup\Pi_{i,j-1}$ within total budget $\gamma^i$.

\noindent \textbf{Intuition.} We discuss our high-level proof strategy for Lemma~\ref{lem:KeyLemmaAux}. 
Our arguments again rely on the notion of ``richness''. Recall from above that $T_{i,j}$ denotes the \Orient profit of tour $\Pi_{i,j}$ at scale $j$.
We call a scale $j$ ``rich'' if $T_{i,j} \geq Y$ and otherwise ``poor''. 
A scale $j$ being poor roughly (but not quite) indicates that $\OPT$ cannot find enough low-cost vertices qualified for scale $j$ outside the tour $\Pi_{i,j}$. 
A scale $j$ being rich implies that $\Pi_{i,j}$ contains enough vertices that are qualified for scale $j$, which is an immediate consequence of Jogdeo-Samuels inequality (Theorem~\ref{thm:JogdeoSamuels}). 
We plan to find a critical scale $\jcrit$ that corresponds to the transition from rich to poor scales.
Notice that such a critical scale $\jcrit$ might be different from the critical scale $\tildejcrit$ found by our algorithm, but we show that it suffices to argue about $\jcrit$ since $\tildejcrit$ is only better. 

To argue about $\jcrit$, we consider the tours corresponding to both scales $\jcrit$ and $\jcrit-1$. Roughly we use the poor scale $\jcrit$ to argue that $\OPT$ cannot find enough low-cost vertices outside these tours and we use the rich scale $\jcrit-1$ to argue that we have enough replacements for these low-cost vertices without paying too much cost. 
Our final analysis is a case analysis depending on whether the transition ever happens or not.
The proof here is more involved than that in Section~\ref{sec:StochKTSP} as we also need to take into account the amount of \Orient profit outside the $C=6000$ repetitions of $\ALGori$ for each scale $j$.

\begin{proofof}{Lemma~\ref{lem:KeyLemmaAux}}
Recall that for any phase $i \geq 0$ and scale $j \in \{0,\cdots,\ell_i\}$, we defined $Y_{i,j}$ to be the maximum number of vertices that can be selected from $\Pi_{i,j} \cup \Pi_{i,j-1}$ with probability at least $0.2$ within cost budget $3\gamma^i$.
We show in the following that $p^*_{i,Y}(\sigma_{i-1}) \geq 0.9$ implies there exists a ``critical'' scale $\jcrit \in \{0,\cdots,\ell_i\}$ with $Y_{i,\jcrit} \geq Y$. 
This critical scale $\jcrit$ might be different from the critical scale $\tildejcrit$ identified by $\ALGStochKCost$.
But since $\tildejcrit$ maximizes $\tildeY_{i,j}$ among all scales $j \in \{0,\cdots,\ell_i\}$, it follows from property (1) in Lemma~\ref{lem:DPestimate} that  $\tildeY_{i,\tildejcrit} \geq \tildeY_{i,\jcrit} \geq Y_{i,\jcrit} \geq Y$.
Now using property (2) in Lemma~\ref{lem:DPestimate} we have that $\tildeY_{i,\tildejcrit} \geq Y$ vertices can be selected from $\Pi_{i,\tildejcrit} \cup \Pi_{i,\tildejcrit-1}$ within cost budget $6\gamma^i$ with probability at least $0.2$. 
It follows that $p_{i,Y}(\sigma_{i-1}) \geq 0.2$.
Therefore, the existence of a critical scale $\jcrit$ with $Y_{i,\jcrit} \geq Y$ would imply Lemma~\ref{lem:KeyLemmaAux}.

In the following, we consider three different cases and prove the existence of such a critical scale $\jcrit$ with $Y_{i,\jcrit} \geq Y$ in each case. 

\noindent \textbf{Case (1): (Scale 0 is poor)}
$T_{i,0} < Y$. 
We show in this case that $p^*_{i,Y}(\sigma_{i-1}) \geq 0.9$ implies that scale $0$ is a ``critical'' scale with $Y_{i,0} \geq Y$. 
Notice that any vertex $v \in V_i$ not qualified for scale $0$ has cost $C_v > \gamma^i$. Therefore, $\OPT$ cannot select any vertex that is not qualified for scale $0$ within total budget $\gamma^i$.

We start by showing that $T_{i,0}^* \leq 0.1$. To prove this, we assume for the purpose of contradiction that $T_{i,0}^* > 0.1$. 
It follow from Lemma~\ref{lem:SmallT*General} that 
\begin{align}
\label{eqn:CostCase1OPTPoor}
\p\Big[\left | \left\{v \in \overline{\Pi}_{i,0}(\OPT): C_v \leq \gamma^i \right\} \right | \leq 200 T_{i,0}^* \Big] ~\geq~ 0.9.
\end{align} 
From Lemma~\ref{lem:LargeT*} we have that
\[
\sum_{v \in  \Pi_{i,0} \setminus \Pi_{i,0}(\OPT)} \p[C_v \leq \gamma^i] \quad \geq \quad  (6000-1) \cdot (T_{i,0}^* - \epsilon) - \epsilon \quad \geq \quad 5000 T_{i,0}^*.
\]
So it follows from Chernoff bound (Theorem~\ref{thm:ChernoffBound}) that
\begin{align}
\label{eqn:CostCase1RicherThanOPT}
\p\Big[ \left | \left\{v \in \Pi_{i,0} \setminus \Pi_{i,0}(\OPT): C_v \leq \gamma^i \right\} \right | \geq 500 T_{i,0}^* \Big] ~\geq~ 0.9.
\end{align}
Since $T_{i,0} < Y$, from Theorem~\ref{thm:JogdeoSamuels} we have that
\begin{align}
\label{eqn:CostCase1WePoor}
\p\Big[ \left | \left\{v \in \Pi_{i,0}: C_v \leq \gamma^i \right\} \right | \leq Y \Big] ~\geq~ 0.5.
\end{align}
Now we count the number of vertices found by $\OPT$ that are qualified for scale $0$ within total budget $\gamma^i$.
It follows from union bound that with probability at least 0.2, all three events in~(\ref{eqn:CostCase1OPTPoor}),~(\ref{eqn:CostCase1RicherThanOPT}) and~(\ref{eqn:CostCase1WePoor}) hold, in which case $\OPT$ finds at most $Y - 300 T_{i,0}^*$ vertices qualified for scale $0$ within total budget $\gamma^i$. Therefore, in order to select at least $Y$ vertices, $\OPT$ needs to select vertices that are not qualified for scale $0$ within total budget $\gamma^i$, which is a contradiction to the assumption that $p^*_{i,Y}(\sigma_{i-1}) \geq 0.9$. 

Therefore we must have $T_{i,0}^* \leq 0.1$. Applying Markov's inequality, the probability that $\OPT$ finds any vertex qualified for scale $0$ in $V_i \setminus \Pi_{i,0}$ is upper bounded by $T_{i,0}^* \leq 0.1$. When this happens and when $\OPT$ selects $Y$ vertices within total budget $\gamma^i$, all vertices selected by $\OPT$ are from $\Pi_{i,0}$. Therefore, with probability at least $p^*_{i,Y}(\sigma_{i-1}) - 0.1 \geq 0.8$, we can select $Y$ vertices from $\Pi_{i,0}$ within cost budget $\gamma^i$ which implies that $Y_{i,0} \geq Y$. 

\smallskip
\noindent \textbf{Case (2): (All scales are rich)} $T_{i,j} \geq Y$ for every scale $j= 0,\cdots, \ell_i$. In this case we show that $\ell_i$ is a ``critical'' scale with $Y_{i,\ell_i} \geq Y$.
Notice that selecting any $Y$ vertices qualified for scale $\ell_i$ has cost at most $Y \cdot \gamma^i/2^{\ell_i} \leq 2 \leq 6\gamma^i$.
Since $T_{i,\ell_i} \geq Y$, it follows from Theorem~\ref{thm:JogdeoSamuels} that with probability no less than $0.5$, at least $Y$ vertices in $\Pi_{i,\ell_i}$ are qualified for scale $\ell_i$ (notice we don't even need the tour $\Pi_{i,\ell_i-1}$ in this case). This implies that $Y_{i,\ell_i} \geq Y$.

\smallskip
\noindent \textbf{Case (3): (Transition from rich to poor scale at $\jcrit$)} $T_{i,0} \geq Y$ but $T_{i,j} < Y$ for some $j \in [\ell_i]$. In this case, let $\jcrit = \arg \min_j \left\{j \in [\ell_i]: T_{i,j} < Y \right\}$. We show in the following that  $\jcrit$ is a ``critical'' scale with $Y_{i,\jcrit} \geq Y$. To prove this, we show that the following two events happen with constant probability:
\begin{enumerate}[topsep=0cm,itemsep=0cm]
\item $T_{i,\jcrit}<Y$ implies that $\OPT$ doesn't find enough vertices {\em qualified} for scale $\jcrit$. This gives a lower bound on the cost of the set of vertices selected by $\OPT$ within total budget $\gamma^i$. 

\item $T_{i,\jcrit-1}\geq Y$ implies that \ALGStochKCost finds enough vertices {\em qualified} for scale $\jcrit-1$. This can be used to upper bound the cost of \ALGStochKCost.
\end{enumerate}

We first consider the sub-case where $T_{i,\jcrit}^* \leq 0.1$. 
This is the case where not many vertices qualified for scale $\jcrit$ can be found outside $\Pi_{i,\jcrit} \cup \Pi_{i,\jcrit-1}$.
In this case, we have $\sum_{v \in \overline{\Pi}_{i,\jcrit}(\OPT)} \p[C_v \leq \gamma^i/2^\jcrit] \leq T_{i,\jcrit}^*$. It follows from Markov's inequality that with probability at least 0.9, $\OPT$ finds no vertex in $V_i \setminus (\Pi_{i,\jcrit} \cup \Pi_{i,\jcrit-1})$ that is qualified for scale $\jcrit$, in which case any vertex $v \in V_i \setminus (\Pi_{i,\jcrit} \cup \Pi_{i,\jcrit-1})$ selected by $\OPT$ has cost $C_v > \gamma^i / 2^\jcrit$. 
Since $T_{i,\jcrit-1} \geq Y$, it follows from Theorem~\ref{thm:JogdeoSamuels} that with probability at least 0.5, we have $\left|\left\{v \in \Pi_{i,\jcrit} \cup \Pi_{i,\jcrit-1}: C_v \leq \gamma^i/2^{\jcrit-1} \right\} \right| \geq Y$.
Furthermore, $p^*_Y \geq 0.9$ implies that with probability at least 0.9, $\OPT$ selects $Y$ vertices within total budget $\gamma^i$.
It follows from union bound that all three events above happen with probability at least 0.2, in which case we can select $Y$ vertices from $\Pi_{i,\jcrit} \cup \Pi_{i,\jcrit-1}$ within cost budget $2\gamma^i$. This implies that $Y_{i,\jcrit} \geq Y$.

Now we deal with the other sub-case where $T_{i,\jcrit}^* > 0.1$. 
This represents the situation where many vertices qualified for scale $\jcrit$ can be found outside $\Pi_{i,\jcrit} \cup \Pi_{i,\jcrit-1}$. 
Roughly, Lemma~\ref{lem:SmallT*General} implies in this case that $\OPT$ finds at most $200T_{i,\jcrit}^*$ low-cost vertices in $V_i \setminus (\Pi_{i,\jcrit} \cup \Pi_{i,\jcrit-1})$.
In general, these vertices might have very tiny cost. 
However, we show in the following Claim~\ref{claim:auxclaim} that increasing the cost of each one of these low-cost vertices to $\gamma^i/2^{\jcrit}$ will increase their total cost by at most $O(\gamma^i)$.
This allows us to lower bound the cost of the set of vertices selected by $\OPT$ within total budget $\gamma^i$.

\begin{claim}
\label{claim:auxclaim}
We have $200 T_{i,\jcrit}^* \cdot \gamma^i / 2^\jcrit \leq \gamma^i/2$. 
\end{claim}

Before proving Claim~\ref{claim:auxclaim}, we  complete the proof of Lemma~\ref{lem:KeyLemmaAux}. 
Since $T_{i,\jcrit}^* > 0.1$, from Lemma~\ref{lem:SmallT*General} we have 
\begin{align}
\label{eqn:CostCase3OPTPoor}
\p\Big[\left | \left\{v \in \overline{\Pi}_{i,\jcrit}(\OPT): C_v \leq \gamma^i/2^\jcrit \right\} \right | \leq 200 T_{i,\jcrit}^* \Big] ~\geq~ 0.9.
\end{align} 
Since $T_{i,\jcrit-1} \geq Y$, it follows from Theorem~\ref{eqn:ALGStochKCostAssume} that
\begin{align}
\label{eqn:CostCase3WeRich}
\p\Big[ \left | \left\{v \in \Pi_{i,\jcrit} \cup \Pi_{i,\jcrit-1}: C_v \leq \gamma^i/2^{\jcrit-1} \right\} \right | \geq Y \Big] ~\geq~ 0.5.
\end{align}
Together with the assumption that $p^*_{i,Y} \geq 0.9$, we have from union bound that with probability at least $0.2$, both events in~(\ref{eqn:CostCase3OPTPoor}) and~(\ref{eqn:CostCase3WeRich}) hold and that $\OPT$ selects at least $Y$ vertices within total budget $\gamma^i$.
When all three events happen, we can replace $Y - |\Pi_{i,\jcrit}(\OPT)|$ vertices in $\overline{\Pi}_{i,\jcrit}(\OPT)$ by $Y - |\Pi_{i,\jcrit}(\OPT)|$ vertices in $(\Pi_{i,\jcrit}
\cup \Pi_{i,\jcrit-1}) \setminus \Pi_{i,\jcrit}(\OPT)$ that are qualified for scale $\jcrit-1$.
Claim~\ref{claim:auxclaim} together with~(\ref{eqn:CostCase3OPTPoor}) imply that after such replacements, we reach a subset of $Y$ vertices in $\Pi_{i,\jcrit} \cup \Pi_{i,\jcrit-1}$ with total cost at most $2(\gamma^i + \gamma^i/2) = 3 \gamma^i$. 
So we conclude that with probability at least $0.2$, we can select at least $Y$ vertices from $\Pi_{i,\jcrit}\cup \Pi_{i,\jcrit-1}$ within cost budget $3\gamma^i$. This implies that $Y_{i,\jcrit} \geq Y$ and finishes the proof of Lemma~\ref{lem:KeyLemmaAux}. 
\end{proofof}

Now we are left to prove Claim~\ref{claim:auxclaim}.

\begin{proofof}{Claim~\ref{claim:auxclaim}}
We consider the tour $\Pi_{i,\jcrit}$.
Denote respectively $\Pi_{i,\jcrit}'(\OPT) \subseteq \Pi_{i,\jcrit}$ and $\overline{\Pi}_{i,\jcrit}'(\OPT) \subseteq V_i \setminus \Pi_{i,\jcrit}$ the (random) set of vertices visited by $\OPT$ inside and outside $\Pi_{i,\jcrit}$ within total budget $\gamma^i$.
Since $T_{i,\jcrit} < Y$, it follows from Theorem~\ref{thm:JogdeoSamuels} that 
\begin{align}
\label{eqn:CostAuxClaimWePoor}
\p\Big[ \left | \left\{v \in \Pi_{i,\jcrit}: C_v \leq \gamma^i/2^\jcrit \right\} \right | \leq Y \Big] ~\geq~ 0.5.
\end{align}
Since $T_{i,\jcrit}^* > 0.1$, Lemma~\ref{lem:SmallT*General} gives
\begin{align}
\label{eqn:CostAuxClaimOPTPoor}
\p\Big[\left | \left\{v \in \overline{\Pi}_{i,\jcrit}'(\OPT): C_v \leq \gamma^i/2^\jcrit \right\} \right | \leq 200 T_{i,\jcrit}^* \Big] ~\geq~ 0.9.
\end{align} 
Lemma~\ref{lem:LargeT*} followed by Chernoff bound gives
\begin{align}
\label{eqn:CostAuxClaimRicherThanOPT}
\p\Big[ \left | \left\{v \in \Pi_{i,\jcrit} \setminus \Pi_{i,\jcrit}'(\OPT): C_v \leq \gamma^i/2^\jcrit \right\} \right | \geq 600T_{i,\jcrit}^* \Big] ~\geq~ 0.9.
\end{align}
From the assumption that $p^*_{i,Y} \geq 0.9$, we have
\begin{align}
\label{eqn:CostAuxClaimOPTDone}
\p\Big[\left\{\text{\OPT selects at least $Y$ vertices within total budget $\gamma^i$}\right\} \Big] ~\geq~ 0.9.  
\end{align}
By union bound, all four events in~(\ref{eqn:CostAuxClaimWePoor})-(\ref{eqn:CostAuxClaimOPTDone}) happen with positive probability, in which case $\OPT$ selects at least $400 T_{i,\jcrit}^*$ vertices that are not qualified for scale $\jcrit$ within total budget $\gamma^i$.
This implies that $400 T_{i,\jcrit}^* \cdot \gamma^i/2^\jcrit \leq \gamma^i$ from which Claim~\ref{claim:auxclaim} immediately follows.
\end{proofof}

\subsection{The Dynamic Programming Sub-procedure $\ALGDP$}
\label{subsec:ALGDP}

In this section, we give our dynamic program \ALGDP and prove Lemma~\ref{lem:DPestimate} restated below for convenience. 

\DPestimate*

Recall, our dynamic program $\ALGDP$ is used to approximately compute the maximum number of vertices that can be selected from a certain tour with probability at least $0.2$ within cost budget $3\gamma^i$.
Essentially, $\ALGDP$ solves the following sub-problem: We are given a target $T \in \mathbb{Z}_{\geq 0}$, a budget $B \geq 0$ and $n$ independent non-negative stochastic costs. The goal is to find  the probability $P_{T,B}$ that there exists a subset $S$ of size $T$ with sum of its costs at most $B$. Since this general problem is NP-hard,  we give a dynamic program that finds something between $P_{T,B}$ and $P_{T,2B}$. 
Lemma~\ref{lem:DPestimate} follows immediately from the following Lemma~\ref{lem:dp}.

\begin{restatable}{lemma}{LemmaDP}
\label{lem:dp}
Given $n$ independent non-negative random variables $V = \{C_1,C_2,...,C_n \}$, a target $T \in \mathbb{Z}_{\geq 0}$ and a budget $B \geq 0$. Let $P_{T,B}$ denote the probability that there exists a subset $S\subseteq V$ of size at least $T$  and $\sum_{i\in S} C_i \leq B$. 
Then there's an efficient dynamic programming $\ALGDP (T,B,V)$ that outputs a value $\widetilde{P}_{T,B}$ s.t. $P_{T,B}\leq \widetilde{P}_{T,B} \leq P_{T,2B}$.
\end{restatable}

\begin{proofof}{Lemma\ref{lem:DPestimate}}
We first prove property (1).
Recall that $Y_{i,j}$ is the maximum number of vertices that can be selected from $\Pi_{i,j} \cup \Pi_{i,j-1}$ with probability at least $0.2$ within cost budget $3\gamma^i$, and
$\tildeY_{i,j}$ is the largest integer such that $\ALGDP(\tildeY_{i,j}, 3\gamma^i, \Pi_{i,j}\cup \Pi_{i,j-1}) \geq 0.2$.
Consider the set of stochastic costs in $\Pi_{i,j} \cup \Pi_{i,j-1}$. 
By definition, we have $P_{Y_{i,j},3\gamma^i} \geq 0.2$.
It follows from Lemma~\ref{lem:dp} that $\ALGDP(Y_{i,j},3\gamma^i,\Pi_{i,j} \cup \Pi_{i,j-1}) \geq 0.2$. 
This implies that $\tildeY_{i,j} \geq Y_{i,j}$ and proves property (1).

Now we prove property (2). Applying Lemma~\ref{lem:dp} we have that $P_{\tildeY_{i,j},6\gamma^i} \geq \ALGDP(\tildeY_{i,j}, 3\gamma^i, \Pi_{i,j}\cup \Pi_{i,j-1}) \geq 0.2$. 
It follows that $\tildeY_{i,j}$ vertices can be selected from $\Pi_{i,j} \cup \Pi_{i,j-1}$ within cost budget $6\gamma^i$ with probability at least $0.2$. 
This establishes property (2) and finishes the proof of Lemma~\ref{lem:DPestimate}.
\end{proofof}

\IGNORE{The following corollary follows immediately from the previous Lemma~\ref{lem:dp}.

\begin{corollary}
\label{cor:dp}
Given $n$ independent non-negative random variables $V=\{C_1,C_2,...,C_n\}$, a  target $T \in \mathbb{Z}^+$, and a budget $B \geq 0$. With probability at least $P_{T,B}^*$, we can find a subset $S \subseteq V$ s.t. $|S| = T$ and  $\sum_{i\in S} C_i\leq 2B$.
\end{corollary}}

\begin{proofof}
{Lemma~\ref{lem:dp}}
We begin by discretizing each $C_i$ to be $\overline{C}_i := \lfloor C_i \cdot n/B \rfloor \in \mathbb{N}$ and define $\overline{P}_{T,n}$ the probability that there exists a subset $S \subseteq V$ s.t. $|S| \geq T$ and $\sum_{i \in S} \overline{C}_i \leq n$. 
Notice that $\sum_{i \in S} \overline{C}_i \leq n$ implies that $\sum_{i \in S} C_i \leq 2B$.
Therefore we have $P_{T,B} \leq \overline{P}_{T,n} \leq P_{T,2B}$.
In the following, we give a dynamic programming $\ALGDP$ that computes the value $\overline{P}_{T,n}$ and we will set $\widetilde{P}_{T,B}$ in the statement of the lemma to be $\overline{P}_{T,n}$.
Assume without loss of generality that $\overline{C}_i \leq n+1$ as one can truncate the distribution of $\overline{C}_i$ at $(n+1)$ without changing $\overline{P}_{T,n}$.

Denote $A(i,j)$ the $j$th smallest value among the first $i$ random variables. We build a DP table where each entry $P[i,j,\ell,m]$ (for $i \in \{1,\cdots,n\}, j \in \{1,\cdots, T\}$, $\ell \in \{0,\cdots, n\}$ and $m \in \{0,\cdots, n\}$) denotes the probability that the smallest $j$ values among the first $i$ random variables sum up to $\ell$ and the $j$th smallest value among the first $i$ random variables is equal to $m$, i.e.    $\sum_{s=1}^j A[i,s]=\ell$ and $A[i,j] =m$.

\smallskip 
\noindent \textbf{Initial values:}  
$\ALGDP$ initializes certain entries of the DP table as follows. 
\begin{itemize}
    \item \textbf{Case 1 (impossible events):} set $P[i,j,\ell,m]$ to be 0 if $j>i$, $m > \ell$ or $m\cdot j < \ell$.
    \item \textbf{Case 2 ($j=1$):} set $P[i,1,\ell,m] =\prod_{s \in [i]} \p[\overline{C}_s \geq \ell] - \prod_{s \in [i]} \p[\overline{C}_s > \ell]$ if $\ell = m$, and 0 otherwise.
\end{itemize} 
Note that all the entries corresponding to $i=1$ are already included in the two cases above. 

\smallskip
\noindent \textbf{Recursion:} $\ALGDP$ uses the following recursion.
\begin{align}
\label{eqn:DPrecursion}
P[i,j,\ell,m] &= \p[\overline{C_i}>m] \cdot P[i-1,j,\ell,m] + \sum_{u=0}^{m-1}\p[\overline{C_i}=u] \cdot P[i-1,j-1,\ell-u,m] \nonumber \\
& \qquad +\p[\overline{C_i}=m] \cdot \Big(\sum_{u=0}^mP[i-1,j-1,\ell-m,m-u]-\sum_{u=1}^mP[i-1,j,\ell-u,m-u] \Big).
\end{align}

\smallskip
\noindent \textbf{Output:} after computing all the entries of the DP table, $\ALGDP$ outputs $\overline{P}_{T,n} = \sum_{\ell=0}^n \sum_{m=0}^\ell P[n,T,\ell,m]$.

Now we prove the correctness of $\ALGDP$. Given the definition of $P[i,j,\ell,m]$, we can immediately verify that the assignment of initial values and the final output are correct if all the entries of the DP table computed from~(\ref{eqn:DPrecursion}) are also correct. 
To see the correctness of the recursion, we consider the outcome of $\overline{C}_i$. 
When $\overline{C}_i > m$, in order to satisfy $\sum_{s=1}^j A[i,s] = \ell$ and $A[i,j] = m$, one must have that $\overline{C}_i$ is not in the $j$ smallest values among the first $i$ random variables. 
This verifies the first term in~(\ref{eqn:DPrecursion}).
When $\overline{C}_i = u < m$, in order to satisfy $\sum_{s=1}^j A[i,s] = \ell$ and $A[i,j] = m$, one must have that $\overline{C}_i$ is one of the $j$th smallest values among the first $i$ random variables. 
Also notice that in this case, the $(j-1)$th smallest value among the first $(i-1)$ random variables is still $m$ and that $\sum_{s=1}^{j-1} A[i-1,s] = \ell-\overline{C}_i$. 
This gives the second term in~(\ref{eqn:DPrecursion}). 

Now we verify the last term in~(\ref{eqn:DPrecursion}, which corresponds to the case where $\overline{C}_i = m$.
In this case, we might as well select $\overline{C}_i$ as one of the $j$ smallest values among the first $i$ random variables.
In order to satisfy $\sum_{s=1}^j A[i,s] = \ell$ and $A[i,j] = m$, we need the smallest $(j-1)$ values among the first $(i-1)$ random variables to sum up to $\ell-m$ and the $(j-1)$th smallest value to be at most $m$, i.e. $\sum_{s=1}^{j-1} A[i,s] = \ell - m$ and $A[i,j-1] \leq m$. 
The probability of this event is exactly $\sum_{u=0}^m P[i-1,j-1,\ell-m,m-u]$. 
However, in order to ensure that $A[i,j]=m$, we also need the $j$th smallest value among the first $(i-1)$ random variables to be at least $m$ (i.e. $A[i-1,j] \geq m$) and the outcomes that don't satisfy this condition needs to be excluded from the previous event.
Putting everything together, we have the following:
\begin{align*}
& \p\Big\{\sum_{s=1}^{j-1}A[i-1,s]=\ell - m, A[i-1,j-1] \leq m, A[i-1,j]\geq m\Big\} \\
&=\p\Big\{\sum_{s=1}^{j-1}A[i-1,s]=\ell - m, A[i-1,j-1]\leq m\Big\} \\ &\qquad -\p\Big\{\sum_{s=1}^{j-1}A[i-1,s]=\ell - m, A[i-1,j-1]\leq m, A[i-1,j]< m\Big\} \\
&=\sum_{u=0}^mP[i-1,j-1,\ell - m,m-u]-\sum_{u=1}^mP[i-1,j,\ell - u,m-u].
\end{align*}
This immediately gives the last term in~(\ref{eqn:DPrecursion}) and finishes the proof of Lemma~\ref{lem:dp}.
\end{proofof}




\paragraph{Acknowledgments}
We are thankful to the anonymous reviewers of ITCS 2020 for many helpful comments on the presentation of this paper.
Jian Li and Daogao Liu are supported in part by the National Natural Science Foundation of China Grant 61822203, 61772297, 61632016, 61761146003, and the Zhongguancun Haihua Institute for Frontier Information Technology and Turing AI Institute of Nanjing.


\appendix

\section{Missing Proofs in Section~\ref{sec:preliminary}}

\ALGOrient*

\begin{proofof}{Lemma~\ref{lem:ALGOrient}}
Assume without loss of generality that $\OPTO > 0$.
Denote $\rho = 3$ the approximation factor of  $\KTSP$ algorithm \ALGkTSP from~\cite{BlumRV-STOC96}. Denote $R_{\max}:= \max_{v \in V} R_v$ and $R_{\min} := \min_{v \in V} R_v$ the maximum and minimum profit in the \Orient instance. 
Notice, $\OPTO \in [R_{\min}, n \cdot R_{\max}]$.

\ALGori applies binary search in $[R_{\min}, n \cdot R_{\max}]$,
starting with profit target $(R_{\min} + n \cdot R_{\max})/2$. 
For each profit target $\lambda$, $\ALGori$ runs $\ALGkTSP$ with target reward $\lambda$ to obtain a tour $\Pi_{\lambda}$ whose length is denoted as $\ell(\Pi_{\lambda})$.
$\ALGori$ performs binary search over $\lambda \in [R_{\min}, n \cdot R_{\max}]$ until finding two values $\lambda_l < \lambda_h \leq \lambda_l + \epsilon$ such that $\ell(\Pi_{\lambda_l}) \leq \rho B$ and $\ell(\Pi_{\lambda_h}) > \rho B$, 
in which case $\ALGori$ returns the tour $\Pi_{\lambda_l}$.
Here we assumed without loss of generality that $\ell(\Pi_{n \cdot R_{\max}}) > \rho B$ as otherwise $\ALGori$ can simply return the tour $\Pi_{n \cdot R_{\max}}$.
Notice that $\ell(\Pi_{\lambda_h}) > \rho B$ implies that $\OPTO < \lambda_h$ and therefore $\ALGori$ finds reward at least $\lambda_l \geq \lambda_h - \epsilon > \OPTO - \epsilon$.
The length of the tour found by $\ALGori$ is $\ell(\Pi_{\lambda_l}) \leq \rho B = O(1) \cdot B$.
This finishes the proof of Lemma~\ref{lem:ALGOrient}.
\end{proofof}

\section{Missing Proofs in Section~\ref{sec:OurApproach}}
\label{sec:MissPfOurApproach}

\KeyLemma*

\begin{proofof}{Lemma~\ref{lem:KeyLemma}}
For any phase $i \geq 0$, denote $u_i$ the probability that $\ALGmeta$ enters phase $i+1$ in the probing stage and $u_i^*$ the probability that $\OPT$ has cost more than $\gamma^i$.
Taking expectation over $\sigma_{i-1}$ for the pre-condition of Lemma~\ref{lem:KeyLemma}, we have $u_i ~\leq~ C \cdot u_i^* + u_{i-1}/\gamma^2$.
It follows that
\[
\sum_{i \geq 1} u_i \cdot \gamma^i ~\leq~ C \cdot \sum_{i \geq 1} u_i^* \cdot \gamma^i + u_0/\gamma + 1/\gamma \cdot \sum_{i \geq 1} u_i \cdot \gamma^i,
\]
which gives
\begin{align}
\label{eqn:KeyLemmaRecur}
(1-1/\gamma) \cdot \sum_{i \geq 1} u_i \cdot \gamma^i ~\leq~ O(1) \cdot \sum_{i \geq 1} u_i^* \cdot \gamma^i + 1/\gamma.
\end{align}
We also notice that 
\[
\OPT \quad \geq \quad  \sum_{i\geq 0}(u_i^*-u_{i+1}^*)\cdot\gamma^i
\quad  =\quad 
(1-1/\gamma)\cdot\sum_{i\geq1}u_i^*\cdot\gamma^i+1,
\]
and that
\[
\ALGmeta
\quad  \leq \quad  O(1)\cdot\sum_{i\geq 0}(u_i-u_{i+1})\cdot\gamma^{i+1} \quad  = \quad  O(1)\cdot\sum_{i\geq 0}u_i\cdot\gamma^i.
\]
It follows from~(\ref{eqn:KeyLemmaRecur}) that $\ALGmeta \leq O(1) \cdot \OPT$. This finishes the proof of Lemma~\ref{lem:KeyLemma}.
\end{proofof}

\section{Missing Proofs in Section~\ref{sec:StochKCost}}
\label{sec:MissPfStochKCost}

\KeyLemStochKCost*

\begin{proofof}{Lemma~\ref{lem:KeyLemStochKCost}}
We fix any outcome $\sigma_{i-1}$ of vertices visited by \ALGStochKCost in the first $i-1$ phases of its probing stage. 
The lemma trivially holds in the case where $u_i^*(\sigma_{i-1}) \geq 0.01$ as we have $100u_i^*(\sigma_{i-1}) \geq 1$. 
If $u_{i-1}(\sigma_{i-1}) = 0$ which means that \ALGStochKCost already selects $k$ vertices before entering phase $i$ in the probing stage, then $u_i(\sigma_{i-1})=0$ and again the lemma trivially holds.
We therefore assume that $u_i^*(\sigma_{i-1}) < 0.01$ and that $u_{i-1}(\sigma_{i-1})=1$.
Now proving Lemma~\ref{lem:KeyLemStochKCost} is equivalent to proving
\begin{align}
\label{eqn:ALGStochKCostAssume}
u_i(\sigma_{i-1}) \leq 100 u_i^*(\sigma_{i-1}) + 1/\gamma^2.
\end{align}

Denote $k(\sigma_{i-1})$ the remaining target at the beginning of phase $i$ in the probing stage of \ALGStochKCost.
We first consider Selection-Process 1 and denote $N_{\old}(\sigma_{i-1})$ the number of vertices selected from $\sigma_{i-1}$ in this process. 
We assume without loss of generality that $N_{\old}(\sigma_{i-1}) < k(\sigma_{i-1})$, as otherwise our algorithm has already selected $k$ vertices after Selection-Process 1 and~(\ref{eqn:ALGStochKCostAssume}) immediately follows.
Since Selection-Process 1 uses cost budget $\gamma^i$ to select as many unselected vertices from $\sigma_{i-1}$ as possible, $\OPT$ can select at most $N_{\old}(\sigma_{i-1}) + k - k(\sigma_{i-1})$ vertices from $\sigma_{i-1}$ within total budget $\gamma^i$. Denote $N_{\new}(\sigma_{i-1}) := k(\sigma_{i-1}) - N_{\old}(\sigma_{i-1})$ the remaining target for $\ALGStochKCost$ after Selection-Process 1. 
It follows that in order to select $k$ vertices within total budget $\gamma^i$, $\OPT$ needs to select at least $N_{\new}(\sigma_{i-1})$ vertices from $V_i$ within total budget $\gamma^i$. Therefore, $u_i^*(\sigma_{i-1}) < 0.01$ implies that $\OPT$ selects at least $N_{\new}(\sigma_{i-1})$ vertices from $V_i$ within total budget $\gamma^i$ with probability at least $0.99$. Applying Lemma~\ref{lem:KeyLemmaAux} with $Y = N_{\new}(\sigma_{i-1})$, it follows that our algorithm finds at least $N_{\new}(\sigma_{i-1})$ vertices from $V_i$ which can be selected within cost budget $6\gamma^i$ with probability at least $0.2$. This implies that $u_i(\sigma_{i-1}) \leq 0.8 \leq 1/\gamma^2$ and~(\ref{eqn:ALGStochKCostAssume}) is established.
\end{proofof}

{\small
\bibliographystyle{alpha}
\bibliography{bib}
}

\end{document}